\documentclass[pdflatex,sn-mathphys-num]{sn-jnl}

\usepackage{graphicx}%
\usepackage{multirow}%
\usepackage{amsmath,amssymb,amsfonts}%
\usepackage{amsthm}%
\usepackage{mathrsfs}%
\usepackage[title]{appendix}%
\usepackage{xcolor}%
\usepackage{textcomp}%
\usepackage{manyfoot}%
\usepackage{booktabs}%
\usepackage{algorithm}%
\usepackage{algorithmicx}%
\usepackage{algpseudocode}%
\usepackage{mhchem}
\usepackage[normalem]{ulem}
\usepackage{listings}%
\RequirePackage{tikz}

\usepackage{xcolor}
\definecolor{SkyBlue}{RGB}{0,206,235}
\definecolor{LimeGreen}{RGB}{50,205,50}

\theoremstyle{thmstyleone}%
\theoremstyle{thmstyletwo}%

\theoremstyle{thmstylethree}%

\begin{document}

\title[Article Title]{ChemXDyn: Dynamics-informed species and reaction detection methodology from atomistic simulations}


\author[1]{\fnm{Raj} \sur{Maddipati}}\email{rajmaddipati@iisc.ac.in}

\author[1]{\fnm{Dhruthi} \sur{Boddapati}}\email{dhruthiv@iisc.ac.in}

\author[2]{\fnm{Elangannan} \sur{Arunan}}\email{arunan@iisc.ac.in}

\author*[1]{\fnm{Phani} \sur{Motamarri}}\email{ phanim@iisc.ac.in}

\author*[1]{\fnm{Konduri} \sur{Aditya}}\email{konduriadi@iisc.ac.in}

\affil[1]{Department of Computational and Data Sciences, Indian Institute of Science, Bengaluru, Karnataka 560012, India}

\affil[2]{Department of Inorganic and Physical Chemistry, Indian Institute of Science, Bengaluru, Karnataka 560012, India}



\abstract{
Accurate identification of chemical species and reaction pathways from molecular dynamics (MD) trajectories is a prerequisite for deriving predictive chemical-kinetic models and for mechanistic discovery in reactive systems. However, state-of-the-art trajectory analysis methods infer bonding from instantaneous distance thresholds, which can misclassify transient, nonreactive encounters as bonds and thereby introduce spurious intermediates, distorted reaction networks, and biased rate estimates. Here, we introduce ChemXDyn, a dynamics-aware computational methodology that leverages time-resolved interatomic distance (IAD) signatures as a core principle to robustly identify chemically consistent bonded interactions and, consequently, extract meaningful reaction pathways. In particular, ChemXDyn propagates molecular connectivity through time while enforcing atomic valence and coordination constraints to distinguish genuine bond-breaking and bond-forming events from transient, nonreactive encounters. We evaluate ChemXDyn on ReaxFF MD simulations of hydrogen and ammonia oxidation and on neural-network potential MD simulations of methane oxidation, and benchmark its performance against widely used trajectory analysis methods. Across these cases, ChemXDyn suppresses unphysical species prevalent in static analyses, recovers experimentally consistent reaction pathways, and improves the fidelity of rate constant estimation. In ammonia oxidation, ChemXDyn removes unphysical intermediates (including \(\ce{N3O}\), \(\ce{N4O}\), \(\ce{N4O2}\), and \(\ce{HN2O2}\)) and resolves key \(\ce{NO_x}\)- and \(\ce{N2O}\)-forming and -consuming routes (for example, $\ce{NH2 + HO2 -> H2NO + OH}$ and $\ce{N2O + H -> N2 + OH}$ ). In methane oxidation, it reconstructs the canonical progression $\ce{CH4 \rightarrow CH3 \rightarrow CH2 \rightarrow CH \rightarrow CHO/CH2O \rightarrow CO \rightarrow CO2}$, which is consistent with established mechanisms yet is often fragmented by threshold-based approaches. By linking atomistic dynamics to chemically consistent reaction identification, ChemXDyn provides a transferable foundation for MD-derived reaction networks and kinetics, with potential utility spanning combustion, heterogeneous catalysis, plasma chemistry, and electrochemical reaction environments.
}




\maketitle

\section*{Introduction}\label{sec1}

A comprehensive understanding of complex reaction pathways is central to advancing technologies that rely on reactive chemistry, including combustion and propulsion, heterogeneous catalysis and reactor design, chemical vapor deposition and materials processing, polymerization, atmospheric chemistry, and electrochemical energy conversion. Driven by sustainability and climate-mitigation goals, there is growing interest in identifying reaction networks and developing predictive chemical kinetic (CK) models for emerging feedstocks and reaction environments, ranging from net-zero fuels and their blends to catalytic and plasma-assisted processes. Such CK descriptions enable systematic screening of operating conditions and chemistries, support mechanism reduction and uncertainty quantification, and provide the mechanistic basis for designing cleaner, more efficient, and robust chemical systems. However, chemical mechanisms often involve hundreds to thousands of species and reactions, which limits direct experimental characterization and complicates model development. Consequently, detailed mechanisms are typically assembled from a combination of theory, experiments, and simulation. While this strategy has been successful for comparatively well-characterized chemistries, the challenge escalates for larger and multifunctional molecules, condensed-phase or interfacial environments, and reactive systems with strong non-equilibrium effects, where comprehensive CK models across wide operating conditions remain very limited.

With advances in computational techniques and the availability of high-performance computing, molecular dynamics (MD) has become a powerful predictive tool that offers detailed, atomistic insights into chemical mechanisms. It serves as a valuable complement to both theoretical models and experimental observations, and has significantly contributed to progress in combustion and energy research, particularly in elucidating reaction pathways from a microscopic perspective \cite{mao2023classical}. In MD simulations, atomic motion follows Newton’s laws, with interatomic forces derived from \textit{ab initio} calculations, classical reactive force fields such as ReaxFF \cite{van2001reaxff, chenoweth2008reaxff, senftle2016reaxff}, or machine learned interatomic potentials (MLIP) \cite{nnmd}. While ReaxFF offers a computationally efficient means to model bond formation and breaking over large systems and long timescales, ML-based potentials provide near-\textit{ab initio} accuracy at a fraction of the cost, enabling chemically consistent simulations across extended spatial and temporal domains. MD simulations generate detailed time-resolved trajectories of atomic positions and velocities. These trajectories are subsequently analyzed to identify  chemical species and reactions across timesteps. The reliability of the inferred chemical mechanisms strongly depends on the accuracy with which molecular species are identified at each timestep.

Traditional analysis tools rely on instantaneous interatomic distances or bond orders with fixed thresholds to infer bonding \cite{agrawalla2011development, dontgen2015automated, brehm2011travis, jcc, pnas}. Such static heuristics, however, are prone to transient bond “flickering” caused by thermal vibrations or transient collisions, leading to spurious species and unphysical reaction events. Recent studies have attempted to incorporate temporal coherence into bonding analysis to overcome this limitation. Rice \textit{et al.}~\cite{rice2020heuristics} introduced a persistence-based criterion that tracks consecutive minima in interatomic distance trajectories (inner turning points) to identify vibrationally sustained bonds, while Hutchings \textit{et al.}~\cite{hutchings2020bond} developed a bond order time-series approach that detects bond formation and dissociation from extrema in the derivative of smoothed bond order curves. On the other hand, ReacNetGenerator \cite{reacnetgenerator} applies hidden Markov models to denoise noisy trajectories before applying conventional threshold-based connectivity analysis. While these approaches represent important steps toward dynamics-aware analysis, they remain empirical, requiring system-specific thresholds for distance, time, or derivative magnitudes and are computationally intensive. None of these methods can accurately determine bond multiplicity (single, double, or triple) or distinguish $\sigma/\pi$ bonding, since they depend solely on time-series trends.

To address limitations in existing MD trajectory analysis, we present ChemXDyn, a dynamics-aware framework that couples time-resolved interatomic distance signatures with chemically informed bond validation to enable reliable identification of species, reaction events, and reaction networks from reactive trajectories. By explicitly incorporating temporal information from trajectories into the analysis, ChemXDyn overcomes the limitations of simple instantaneous distance-based methods, providing improved detection of species and reactions by mitigating spurious bonding, fragmented intermediates, and inflated reaction counts.
We evaluate ChemXDyn on a representative set of reactive systems spanning distinct chemical regimes and simulation paradigms: ReaxFF MD of hydrogen and ammonia oxidation and neural-network potential MD of methane oxidation, and benchmark against widely used trajectory analysis tools. Across these cases, ChemXDyn yields reaction networks that better align with experimentally supported pathways and established mechanisms while substantially reducing unphysical species and reactive events. For example, in ammonia oxidation ChemXDyn identifies 39 unique species and 173 distinct reactions, compared with 68 species and 273 reactions reported by ChemTraYzer for the same trajectory \cite{dontgen2015automated}. ChemXDyn also improves rate constant estimation in hydrogen oxidation, achieving closer agreement with experimental data and reference kinetic models than ChemTraYzer-based pipelines. In methane oxidation, ChemXDyn reconstructs the canonical oxidation cascade from \ce{CH4} to \ce{CO2}, including intermediate radical chemistry (for example, \ce{CH}) that is often missed in static analyses. Collectively, these results establish ChemXDyn as an accurate and transferable computational methodology for converting reactive MD trajectories into chemically consistent mechanisms and kinetic descriptors.

\begin{figure*}[h]
\centering
\includegraphics[width=1\linewidth]{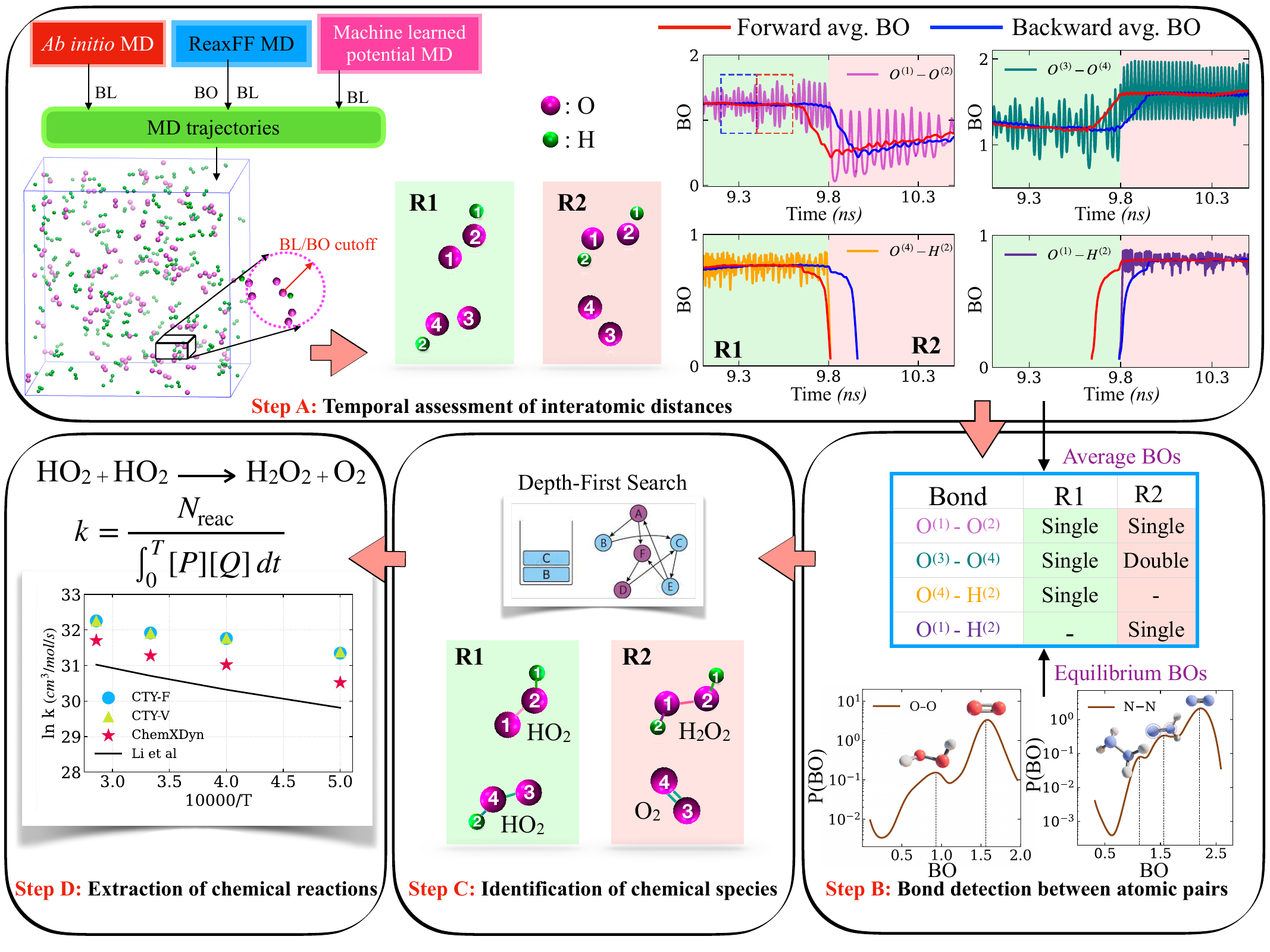}
\caption{ChemXDyn workflow for extracting reactions and rate constants from MD trajectories.
\textbf{Step A:} Two successive MD snapshots (R1 and R2) are analyzed, with bond order (BO) trajectories plotted for four representative bonds. Each trajectory plot includes backward (blue) and forward (red) time-window averages of BO(t), allowing us to distinguish between stable bonds and transient events associated with bond formation, breakage, or transition states.
\textbf{Step B:} Temporally averaged BOs from Step A are compared with equilibrium BO (or IAD) reference values obtained from peaks of BO distribution plots to classify bonds as single, double, etc., while discarding weak, non-bonded interactions.
\textbf{Step C:} The validated bonds and their assigned multiplicities are used to construct a connectivity graph, and a depth-first search (DFS) algorithm groups atoms into molecular species at each timestep.
\textbf{Step D:} Species from consecutive timesteps are compared to detect reactions and further these reaction counts ($\ce{N_{reac}}$) are accumulated across the trajectory to evaluate rate constants as a function of temperature for different systems.}
\label{fig:flowchart}
\end{figure*}

\section*{Results} \label{sec:RESULTS}

\subsection*{Proposed methodology: ChemXDyn} \label{sec:chemxdyn}

Current state-of-the-art trajectory analysis techniques rely on threshold-based criteria using interatomic distances (or related metrics such as bond order) to determine the existence of chemical bonds between atom-pairs, thereby identifying molecular species at each timestep of the simulation. These thresholds are typically heuristic, being either empirically determined \cite{dontgen2015automated, pnas, agrawalla2011development, reacnetgenerator} or derived from the first minima observed in atom-pair distance distribution functions \cite{jcc, qi2012simulations} \textcolor{blue}{(see \textit{SI Appendix A)}}. Although computationally inexpensive, this approach does not incorporate temporal information, making it challenging to accurately distinguish between reactive and non-reactive molecular collisions \textcolor{blue}{(see \textit{SI video V1})}. Furthermore, at elevated temperatures, bond vibrations can cause interatomic distances to transiently fluctuate above or below a interatomic distance threshold for bond existence, leading to the erroneous detection of bond-breaking or bond-forming events occurring at frequencies associated with bond vibrations\textcolor{blue} {(see \textit{SI video V2})}. Such erroneous identification of molecular species at individual timesteps further propagates into inaccuracies in reaction detection, ultimately compromising the accuracy of inferred reactions. To address these issues, the proposed methodology leverages the dynamical information available from MD simulations to significantly enhance the accuracy and reliability of identified species and reactions.

Before proceeding to illustrate our methodology, note that MD simulations produce detailed trajectory data, providing atomic positions and velocities over time. In simulations using reactive force fields, such as ReaxFF, interatomic distances can alternatively be represented by bond orders, which are continuous functions of atomic separation \cite{chenoweth2008reaxff}. By incorporating temporal analysis of interatomic distances, the proposed algorithm systematically analyzes MD trajectories to identify molecular species and chemical reactions through four principal steps, briefly summarized below and described in detail in the subsequent sections. A schematic overview of the ChemXDyn methodology is shown in Fig.~\ref{fig:flowchart}.

\begin{itemize}
    \item[\textbf{Step}] \textbf{A. Temporal assessment of interatomic distances:} Identify atomic clusters by grouping atoms that are within proximity, indicative of a high likelihood of chemical bonding. Compute average interatomic distances for identified atom-pairs within each of these clusters, using a moving time window centered around each timestep to capture its local temporal fluctuations. 
    \item[\textbf{Step}] \textbf{B. Bond detection between atom-pairs:} Assign the bond type (single, double, or triple) between atom-pairs by comparing their average interatomic distances (from Step A) with known equilibrium bond lengths, and validate the assignment by evaluating the coordination number of each atom.
    \item[\textbf{Step}] \textbf{C. Identification of chemical species:} Construct a molecular connectivity graph from bonded atom-pairs (from Step B) and use it to identify distinct chemical species.
    \item[\textbf{Step}] \textbf{D. Extraction of chemical reactions and rate constants:} Identify chemical reactions by comparing chemical species across successive timesteps and further obtain the rate constants.
\end{itemize}

\subsubsection*{Step A. Temporal assessment of interatomic distances}\label{sub:A}
To identify chemical species in a molecular dynamics (MD) system, we begin with an $n$-atom system that yields ${}^{n}C_{2}$ possible atom-pairs. Because most atom-pairs are separated by large interatomic distances (IADs) and have a negligible likelihood of forming bonds, analyzing all possible pairs is both unnecessary and computationally expensive. To address this, we apply a conservative distance-based filtering criterion derived from the IAD distribution across all possible pairs \textcolor{blue}{(see \textit{SI Appendix A})}. This threshold, which is chosen to be significantly larger than any typical covalent bond length or equivalently, smaller than the bond order characteristic of bonded atoms in ReaxFF-based MD, helps isolate atomic clusters with a high likelihood of chemical bonding. This clustering phase is illustrated in Step A of Fig.~\ref{fig:flowchart}.


Next, we analyze the dynamics of atom-pairs within each atomic cluster to accurately detect the presence of chemical bonds. This analysis differentiates true reactive collisions from non-reactive encounters that do not result in bond formation, effectively suppressing spurious bonds and reducing erroneous species identification. This distinction is made possible by exploiting a well-established fact that bonded atom-pairs exhibit characteristic vibrational patterns in their IADs, serving as a reliable indicator of stable bonding, in contrast to the previous methods that rely solely on instantaneous distance-based thresholds \cite{agrawalla2011development, dontgen2015automated, brehm2011travis, pnas, reacnetgenerator}. 


To capture the temporal dynamics of bonding behavior, one can employ time-resolved analyses of IADs using Fourier transforms or obtaining the power spectrum. Assessing vibrational spectra for each atom-pair across all timesteps could, in principle, yield detailed insights into bond persistence and dynamics. However, for large-scale simulations, this becomes intractable, as it requires analyzing every possible atom-pair at each timestep. Instead, we leverage a key insight from spectral analysis: the zeroth-frequency component of the Fourier transform corresponds to the time-averaged IAD for a given atom-pair. For bonded pairs, this average distance is expected to be equal or close to the equilibrium bond length, whereas for unbonded pairs, it tends to be significantly different (higher). To enable this classification, we implement a symmetric moving time window centered at each timestep to calculate both forward and backward window–averaged IADs as a function of time. In particular, incorporating information from both forward and backward windows at each timestep enables accurate detection of bond formation and breakage, which are inherently localized in time, as described subsequently using an example.

Plots in Step A of Fig.~\ref{fig:flowchart} show representative bond order (BO) trajectories obtained from a ReaxFF-based MD simulation of the $\mathrm{H_2/O_2}$ system (Case 1 in Table \ref{tab:1}), where BO acts as a proxy for interatomic distance. The shaded light-green and light-red regions correspond to two consecutive time intervals, $R1$ and $R2$, representing the two different chemical states. The molecular configurations of six tracked atoms are shown for these regimes to illustrate how bond evolution accompanies atomic rearrangements. In the BO trajectory plots, the dashed blue and red rectangles mark the backward and forward temporal windows used for averaging. When the backward (blue line) and forward (red line) window-averaged BOs are nearly identical, the bond is stable in that interval. In contrast, significant deviation between the averages indicates a bond undergoing formation or breakage with time. For example, the $\mathrm{O^{(3)}\!-\!O^{(4)}}$ pair (superscript indicates atom label) exhibits a pronounced divergence between the forward and backward average values during the period of bond rearrangement associated with the $\ce{HO2}$ to $\ce{O2}$ conversion, while the two averages remain nearly identical before and after the transition. A similar divergence is observed for the $\mathrm{O^{(1)}\!-\!O^{(2)}}$ pair, corresponding to the $\ce{HO2}$ to $\ce{H2O2}$ conversion, which is accompanied by concurrent $\mathrm{O^{(1)}\!-\!H^{(2)}}$ bond formation and $\mathrm{O^{(4)}\!-\!H^{(2)}}$ bond dissociation.
These temporal comparisons between forward and backward average BOs provide an objective criterion to distinguish stable bonds from those undergoing reorganization.

In summary, this step in the trajectory analysis identifies atom-pairs with a high likelihood of bonding using temporal variations of IADs within a local time window. A discussion on the choice of time window size and its influence on the results is provided in \textcolor{blue}{\textit{SI Appendix B}}.

\subsubsection*{Step B. Bond detection between atom-pairs}\label{sub:B}

While Step A identifies atom-pairs with a high bonding probability based on temporally averaged IADs (or BOs), this alone does not guarantee the presence of a true chemical bond. Atom-pairs within the same molecule can also exhibit characteristic vibrational patterns even without being directly bonded due to coupling with nearby bonded atoms. For instance, the H–H interaction within a water molecule can exhibit apparent vibrational correlations even in the absence of a direct bond, arising from the bending mode of the $\ce{H2O}$ structure. Accordingly, the objective of this step is to identify atom pairs that are genuinely bonded at each timestep and to classify their bond multiplicities (e.g., single, double, or triple). This is achieved through the procedure outlined below.

First, we assess whether the averaged IADs of each atom-pair lies close to an equilibrium bond length associated with a specific bond type \textcolor{blue}{(see \textit{SI Appendix A})}. These equilibrium bond lengths (or equilibrium bond orders) correspond to the maxima of IAD (or bond order) distribution plots that remain remarkably consistent across different systems and conditions \textcolor{blue}{(see \textit{SI Figure B1})}. These maxima reflect the most probable bonding configurations observed in an MD simulation and correspond to chemically meaningful bond multiplicities. This interpretation aligns with the underlying free energy landscape, where peaks in the distribution correspond to thermodynamically stable states, and minima delineate transitions between bonded and non-bonded regimes. {Step B panel} in Fig~\ref{fig:flowchart} illustrates this behavior for representative cases: the left plot shows the O–O {bond order} distribution from a ReaxFF-based NVT simulation of the $\ce{H_2}/\ce{O_2}$ system at 2000K (Case 1 in Table \ref{tab:1}), while the right plot shows the N–N distribution from a $\ce{NH_3}/\ce{N_2}/\ce{O_2}$ simulation at 1500K (Case 2 in Table \ref{tab:1}). It can be observed that the O–O distribution exhibits two distinct peaks corresponding to single and double bond character, whereas the N–N distribution reveals three well-resolved maxima associated with single, double, and triple bonds. These distributions form the basis for bond multiplicity assignment in the proposed ChemXDyn framework, enabling robust classification of bonded interactions while effectively filtering out unbonded atom-pairs that were spuriously identified in Step A.


Based on the IAD distributions described above, bond multiplicities are determined by comparing the time-averaged IAD of each atom-pair with reference equilibrium values and selecting the closest match. For atom-pairs whose averaged IAD lies near the equilibrium distance of a single bond, an additional weak interaction threshold is applied on the right side of the reference peak (or on the left side for BO) to prevent weak, non-bonded interactions from being misidentified as true chemical bonds. In contrast, higher-order bonds are naturally resolved because their average IAD values lie between adjacent equilibrium maxima, allowing multiplicity to be determined by simple proximity to the nearest equilibrium. When the average IAD fluctuates between two neighboring equilibrium values, the assignment is refined by examining the local bonding environment, as described in the next section. This procedure enables robust identification of chemically meaningful bonds while excluding spurious interactions. Details regarding the choice of the weak interaction threshold, the averaging window size, and their influence on species recognition are provided in \textcolor{blue}{\textit{SI Appendix B}}. At each timestep, accepted bonds are ordered by IAD, and coordination numbers are monitored to enforce valence rules: for example, hydrogen (1), oxygen (2), nitrogen (3), and carbon (4).

In summary, this step processes candidate bonds in order of increasing IADs, prioritizing chemically plausible configurations while systematically excluding weaker, excess interactions that would violate a given atom coordination number. This ensures the construction of physically realistic bonding networks and prevents the formation of unphysical species.

\subsubsection*{Step C. Identification of chemical species}\label{sub:C}
With a validated set of bonded atom-pairs, resulting from Step B established for each timestep, molecular species are identified through the construction of a connectivity graph. In this representation, atoms are treated as nodes and valid bonds that are obtained from Step B, annotated with their assigned multiplicities, serve as edges as shown in Step C panel of Fig.~\ref{fig:flowchart}. A depth-first search (DFS) algorithm is then applied to traverse the graph and group atoms into connected components, each corresponding to a distinct molecular species \cite{dontgen2015automated}. This approach captures the molecular topology at each timestep and distinguishes structural isomers (species with identical atomic compositions but different bonding patterns). The procedure is repeated at every timestep to construct a time-resolved map of species evolution throughout the simulation. 
Each identified species is then converted into a canonical representation, such as a SMILES string, encoding both atomic identities and bonding information. These standardized forms enable consistent species tracking across timesteps and serve as the foundation for reaction detection. 


\subsubsection*{Step D. Extraction of chemical reactions and rate constants}\label{sub:D}

To capture chemical transformations over time, the molecular species identified at consecutive timesteps are compared to detect changes in atomic connectivity. A reaction event is registered whenever one or more bonds form or break, leading to a change in species composition. For each detected event, the corresponding reactant and product species are mapped using their canonical SMILES representations to ensure atom-level correspondence across timesteps. Each unique reaction is then assigned a distinct label, and its cumulative number of occurrences is tracked throughout the trajectory. Step D panel in Fig.~\ref{fig:flowchart} illustrates this procedure for the representative reaction $\ce{HO2 + HO2 -> H2O2 + O2}$, where the formation and breakage of O–O and O–H bonds mark the conversion of $\ce{HO2}$ radicals into $\ce{H2O2}$ and $\ce{O2}$ molecules.
The reaction occurrence counts provide the foundation for rate constant estimation within a time-integrated mass-action framework. For a general bimolecular reaction of the form $P + Q \rightarrow R$, the rate constant $k$ is evaluated as \cite{dontgen2015automated,jcc}:
\begin{equation}
k = \frac{N_{\mathrm{reac}}}{\displaystyle \int_0^T [P][Q]\ dt}
\label{eq:rate_constant}
\end{equation}
Here, $N_{\mathrm{reac}}$ is the total number of reaction events recorded over the simulation time $T$, while $[P]$ and $[Q]$ denote the instantaneous concentrations of reactant species $P$ and $Q$. These concentrations are obtained from species counts at each timestep, normalized by the system volume and converted to mol/cm³ using Avogadro’s number. This formulation naturally extends to unimolecular and higher-order reactions, enabling direct computation of rate constants from atomistic simulations. The rate constants obtained from independent NVT simulations at multiple temperatures can subsequently be used to parameterize rate models such as the Arrhenius equation, as illustrated in the Step D panel of Fig.~\ref{fig:flowchart}.

The ability to accurately extract chemically resolved reaction networks along with their corresponding rate constants directly from MD trajectories using ChemXDyn (Steps A to D) serves as a key enabler for constructing a comprehensive kinetic model that captures both the mechanistic and dynamic aspects of the reactive system. The framework is fully automated, scalable, and broadly applicable to a wide range of reactive molecular dynamics simulations.

\subsection*{ChemXDyn: Performance benchmarks}
To evaluate the performance and generalizability of the ChemXDyn framework, we applied it to reactive molecular dynamics (MD) simulations of three representative fuel-oxidations: hydrogen-oxygen ($\mathrm{H_2}/\mathrm{O_2}$), ammonia-air ($\mathrm{NH_3}/\mathrm{air}$), and methane-oxygen ($\mathrm{CH_4}/\mathrm{O_2}$). Hydrogen, a well-characterized fuel with simple combustion chemistry, serves as an ideal benchmark system. Ammonia has garnered growing interest as a carbon-free energy carrier, yet presents challenges due to its complex nitrogen-based reaction pathways. Methane, a widely studied hydrocarbon is used as a test case that employs neural network-based MLIP in place of ReaxFF for MD simulations, allowing us to assess ChemXDyn's robustness under different force field paradigms. The simulation details associated with the three fuels employed in reactive MD simulations are tabulated in Table~\ref{tab:1}.

\begin{table}
  \centering
  \begin{tabular}{ccccc}
  \hline
    \textbf{Case} &  \textbf{Fuel}  &  \textbf{No. of molecules} & \textbf{interatomic}   & \textbf{Temperature} \\
     &    &  \textbf{Fuel/O$_{2}$/N$_{2}$} & \textbf{potential}   & \textbf{(K)} \\

    \hline
    \hline
    1 & H$_{2}$ & 160/80/0 &  ReaxFF\cite{agrawal}/MLIP\cite{nnmd}  & 2000-3500 \\

    \hline
    2 & NH$_{3}$ & 67/50/195 & ReaxFF\cite{kowalik2019atomistic}  & 2000 \\
    \hline
    
    3 & CH$_{4}$ & 50/90/0 & MLIP\cite{nnmd}  & 3000 \\
  

    \hline
  \end{tabular}
  \caption{Simulation cases considered in this study, detailing the chosen fuel-oxidizer composition (Fuel/O$_{2}$/N$_{2}$), interaction potential employed, and temperature range. }
  \label{tab:1}
\end{table}
The time-resolved atomic trajectories obtained from reactive MD simulations are analyzed to extract the underlying chemical dynamics, either in terms of interatomic distances (IAD) or bond orders (BO) between atom-pairs. Notably, the proposed ChemXDyn framework is designed to flexibly accommodate both descriptors. To benchmark the robustness of our approach, we compare its performance against two state-of-the-art trajectory analyzers, ChemTrYzer \cite{dontgen2015automated} and ReacNetGenerator \cite{reacnetgenerator}.

\subsubsection*{$\mathbf{{H_2}/{O_2}}$ MD trajectory analysis}
\begin{figure*}[h!]
\centering
\includegraphics[width=1\linewidth]{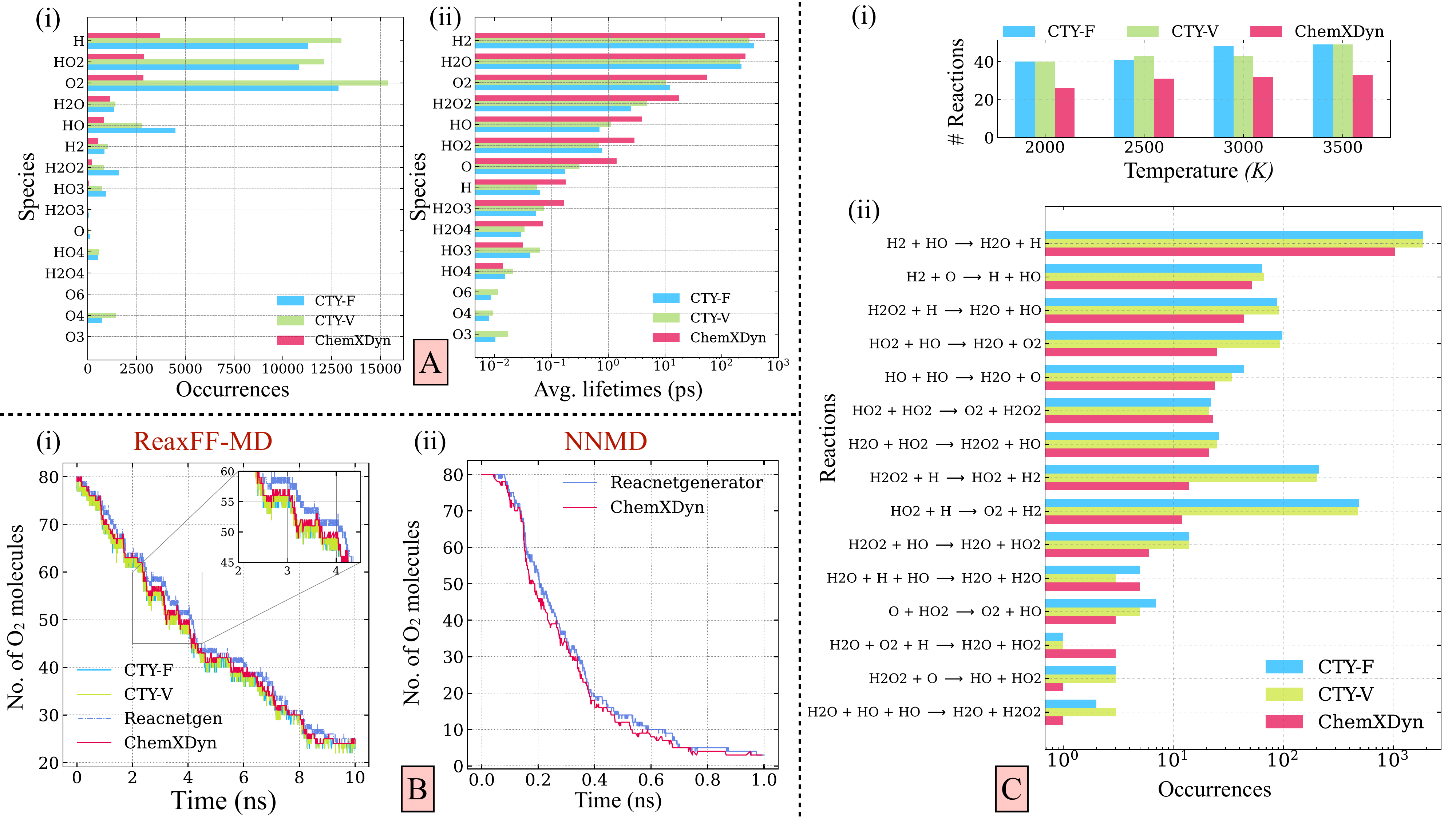}
\caption{Species and reaction detection in $\ce{H2}/\ce{O2}$ oxidation.
(A) Comparison of (i) species occurrences  and (ii) average lifetimes obtained from different trajectory analysis methods.
(B) Temporal evolution of $\ce{O2}$ for (i) ReaxFF MD  and (ii) NNMD  trajectories, highlighting differences between methods.
(C) (i) Total number of reactions detected by each method across temperatures. (ii) Cumulative counts of those reactions present in Li \textit{et al.} mechanism \cite{li} across temperatures obtained using CTY-F, CTY-V and ChemXDyn.}
\label{fig:H2_results}
\end{figure*}
We begin by presenting the results from the successive stages of ChemXDyn, where molecular species are identified in accordance with Steps~A–C of the proposed methodology. As a first outcome, the detected species are characterized by their average lifetimes and number of occurrences. Figure~\ref{fig:H2_results}(A) compares results from three trajectory-analysis methods: (a) ChemTrYzer with a fixed bond order threshold which is typically 0.3 (CTY-F), (b) ChemTrYzer with atom-pair-specific variable thresholds (CTY-V) \cite{jcc}, and (c) the proposed ChemXDyn framework. As shown in panel (i) of Fig.~\ref{fig:H2_results}(A), both CTY-F and CTY-V substantially overestimate the occurrence frequencies of key species ($\ce{H2}$, $\ce{O2}$, $\ce{HO2}$, $\ce{OH}$, $\ce{H}$, $\ce{O}$, $\ce{H2O2}$, $\ce{H2O}$). This inflation, in turn, leads to systematically shorter lifetimes for these species, as evident from panel (ii) of Fig.~\ref{fig:H2_results}(A). This discrepancy is especially pronounced for stable species such as $\ce{O2}$, and is further compounded by the frequent detection of unphysical intermediates, including $\ce{O4}$, and $\ce{O6}$, underscoring fundamental limitations in their bond-detection schemes. Such spurious detections are significantly mitigated using ChemXDyn.

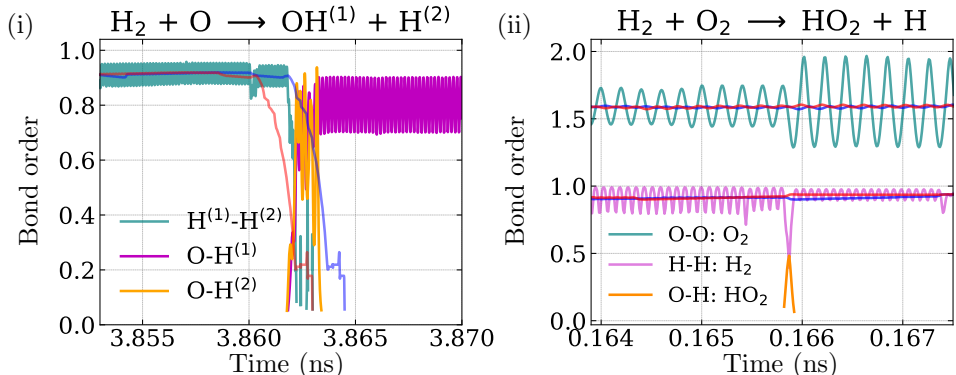
\begin{figure}[!h]
    \centering
  \includegraphics[width=.47\columnwidth]{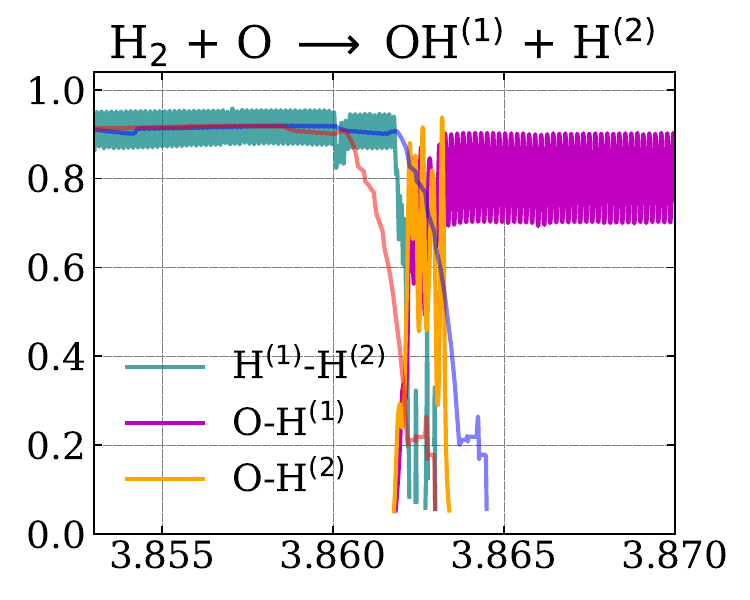}\hspace{2mm}
  \includegraphics[width=.47\columnwidth]{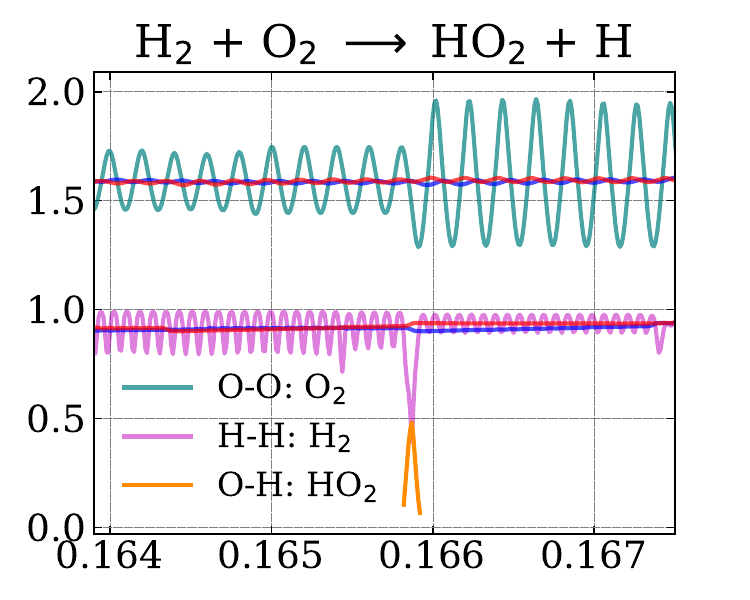}
  \begin{tikzpicture}[remember picture,overlay]
      \node[rotate=90] at (-12.9,2.45) {Bond order};
      \node[rotate=0] at (-9.5,0.0) {Time (ns)};
      \node[rotate=90] at (-6.5,2.45) {Bond order};
      \node[rotate=0] at (-3.0,0.0) {Time (ns)};
      \node[rotate=0] at (-13.0,4.5) {(i)};
      \node[rotate=0] at (-6.5,4.5) {(ii)};

    \end{tikzpicture}

\caption{Temporal bond order (BO) evolution for different atom-pairs involved in two distinct collision events. (i) A genuine reaction event where the $H^{(1)}$–$H^{(2)}$ bond breaks and an $O$–$H^{(1)}$ bond forms, captured by the deviation of both forward (red) and backward (blue) averaged BOs.
(ii) A transient, nonreactive collision between $\ce{H2}$ and $\ce{O2}$ showing no sustained deviation in averaged BOs. A video of this collision event is available in \textcolor{blue}{(see \textit{SI video V1})}.}
  \label{fig:R1_R2}
\end{figure}


We next examine the temporal evolution of $\ce{O2}$ molecules from the ReaxFF MD trajectories, shown in panel (i) of Fig.~\ref{fig:H2_results}(B). Consistent with the inflated occurrence counts and shortened lifetimes discussed above, both CTY-F and CTY-V exhibit pronounced fluctuations in the $\ce{O2}$ time profiles. We also investigated ReacNetGenerator, which, like CTY-F and CTY-V, applies frame-by-frame distance-based thresholds without accounting for temporal variations across frames. This method produces even more inflated and erratic $\ce{O2}$ profiles, further emphasizing the limitations of instantaneous thresholding. In contrast, ChemXDyn provides a smoother and consistent evolution of $\ce{O2}$, characterized by fewer occurrences, longer lifetimes, and mitigated the spurious oscillations. This improvement arises because ChemXDyn incorporates time-window averaging, whereas CTY-F, CTY-V, and ReacNetGenerator neglect the temporal evolution of BOs or IADs. Such frame-wise threshold schemes often generate recurrent artifacts. For example, when two stable $\ce{O2}$ molecules approach each other, transient increases in bond order can surpass the threshold, spuriously forming bonds and producing unphysical species such as $\ce{O4}$ and $\ce{O6}$. Likewise, high temperature vibrations within a stable molecule can transiently elongate bond distances, causing the bond order to drop below the cutoff and triggering artificial dissociation and recombination events (e.g., $\ce{O2 \Leftrightarrow O + O}$). These artifacts collectively inflate species counts and underestimate lifetimes.

To further assess the generality of ChemXDyn, we performed neural network–based molecular dynamics (NNMD) simulations of the $\ce{H2}/\ce{O2}$ system using an MLIP model \cite{nnmd} and analyzed the resulting trajectories with both ChemXDyn and ReacNetGenerator. As shown in panel (ii) of Fig.~\ref{fig:H2_results}(B), ReacNetGenerator again produces inflated $\ce{O2}$ occurrence counts with pronounced temporal oscillations, similar to those observed in the ReaxFF trajectories. In contrast, ChemXDyn yields smooth and chemically consistent $\ce{O2}$ profiles, maintaining physically realistic and stable species evolution across both ReaxFF-MD and NNMD simulations.

\begin{figure*}[h!]
    \centering
  \includegraphics[width=.47\columnwidth]{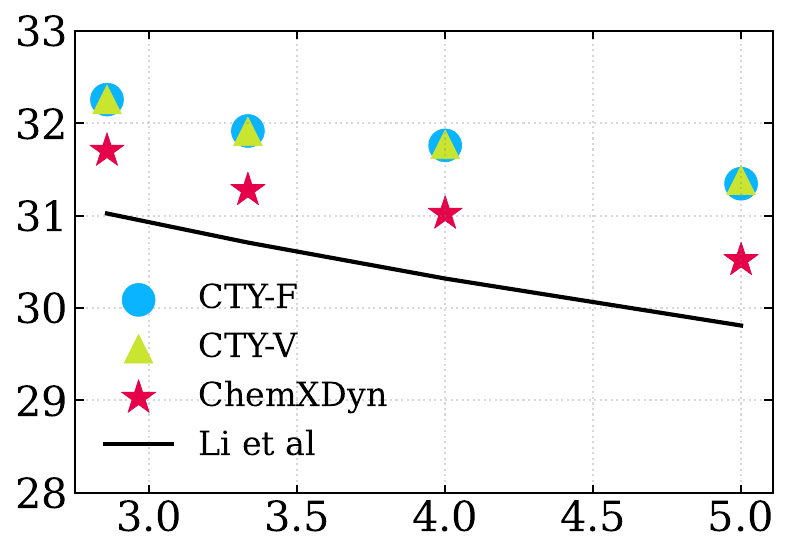}\hspace{3mm}
  \includegraphics[width=.47\columnwidth]{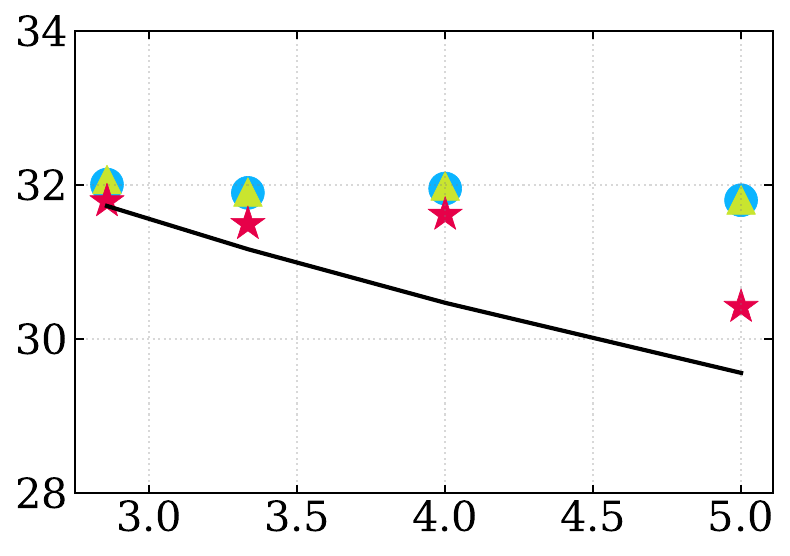}\hspace{3mm}
  
  
  \includegraphics[width=.47\columnwidth]{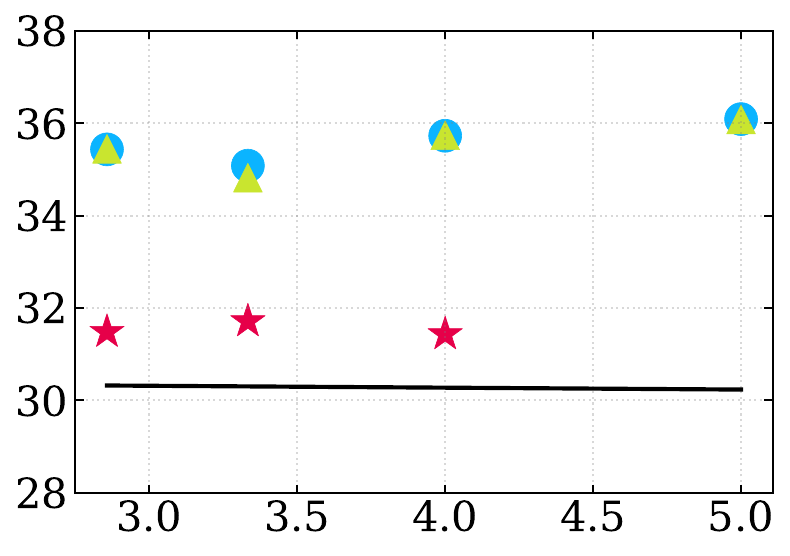}\hspace{3mm}
  \includegraphics[width=.47\columnwidth]{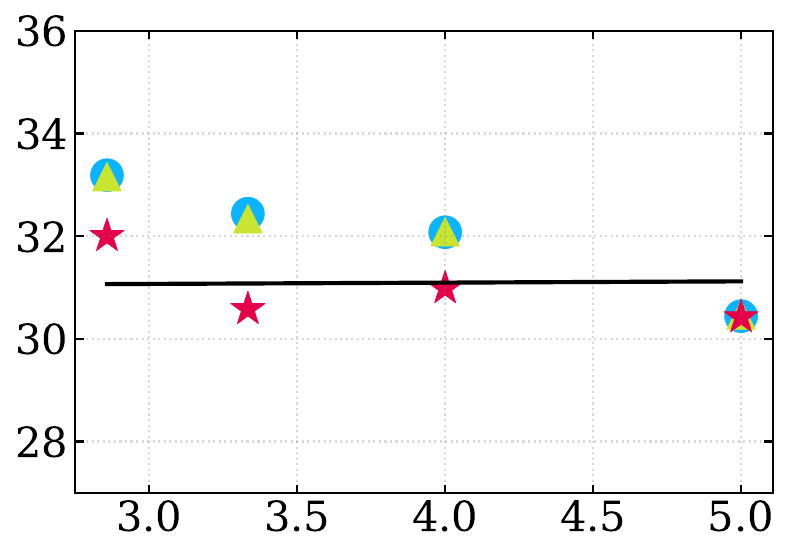}      
  \begin{tikzpicture}[remember picture,overlay]
    \node[rotate=90] at (-13.1,2.4) {ln k \tiny{(cm$^{3}$/mol/s)} };
    \node[rotate=90] at (-13.1,6.6) {ln k \tiny{(cm$^{3}$/mol/s)} };
    \node[rotate=90] at (-6.5,2.4) {ln k \tiny{(cm$^{3}$/mol/s)} };
    \node[rotate=90] at (-6.5,6.6) {ln k \tiny{(cm$^{3}$/mol/s)} };



      \node[rotate=0] at (-9.3,-0.15) {10000/T \tiny{(K$^{-1}$)}};
      \node[rotate=0] at (-3.0,-0.15) {10000/T \tiny{(K$^{-1}$)}}; 

    \node[rotate=0] at (-9.5,8.9) {$\ce{H2 + OH -> H2O + H}$};
    \node[rotate=0] at (-3.2,8.9) {$\ce{H2 + O -> H + OH}$};
    \node[rotate=0] at (-9.5,4.3) {$\ce{HO2 + H -> H2 + O2}$};
    \node[rotate=0] at (-3.2,4.3) {$\ce{HO2 + OH -> H2O + O2}$};
    \end{tikzpicture}

    \vspace{3mm}

  \caption{Rate constants ($k$) for individual reactions obtained from NVT simulations at four temperatures ($T$) using CTY-F, CTY-V and ChemXDyn, compared with the Arrhenius fit obtained from the Li \textit{et al.} mechanism \cite{li}.}
  \label{fig:lnk_plots}
\end{figure*}
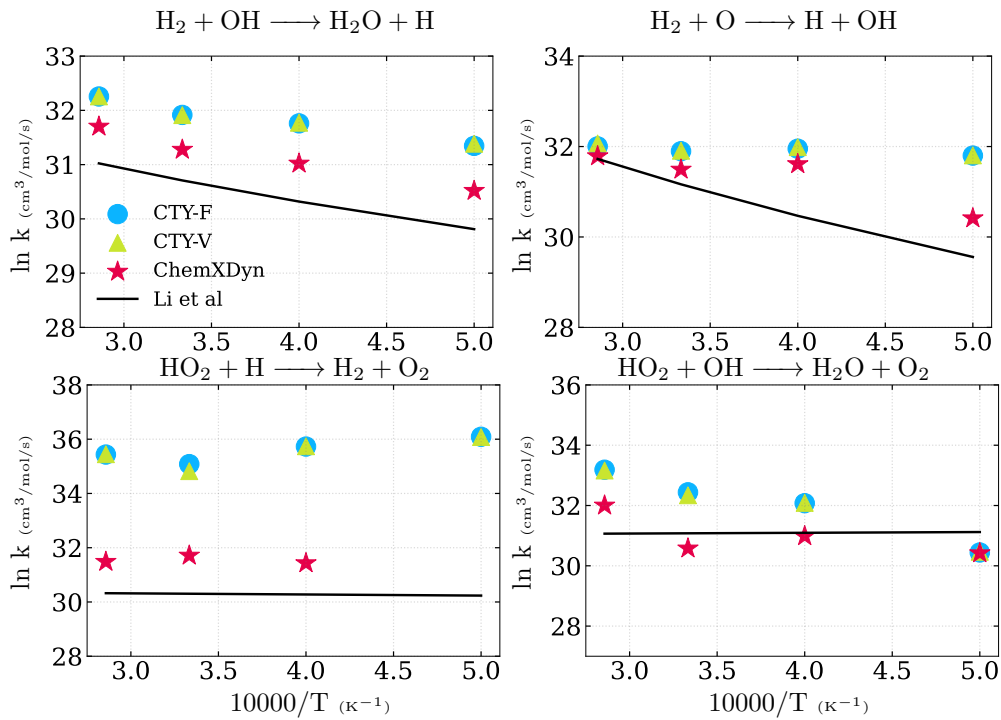

The enhanced accuracy in species identification achieved by ChemXDyn translates directly into more reliable reaction detection. Furthermore, accurate species lifetimes and occurrence profiles are essential for mapping bond rearrangements onto specific chemical transformations and for calculating rate constants from observed reaction frequencies and concentration trajectories. To evaluate this capability, we performed NVT ReaxFF simulations of the $\ce{H2}$/$\ce{O2}$ system at four temperatures (2000 K, 2500 K, 3000 K, and 3500 K) using the ReaxFF potential \cite{agrawalla2011development}. Panel (i) of Fig.~\ref{fig:H2_results}(C) compares the total number of reactions detected by the three methods at each temperature. ChemXDyn consistently reports fewer reactions than CTY-F and CTY-V. To further evaluate the chemical fidelity of these detections, reaction occurrences from each trajectory were aggregated, and the cumulative counts of reactions that are part of the well-established Li \textit{et al.} mechanism \cite{li} for $\ce{H2}$/$\ce{O2}$ combustion were compared across methods (panel (ii) of Fig.~\ref{fig:H2_results}(C)). The differences are prominent: CTY-F and CTY-V substantially over-count reactions because they cannot reliably distinguish reactive events from non-reactive collisions. Their reliance on static bond order thresholds applied frame by frame leads to frequent misclassification of transient contacts as reactions. In contrast, ChemXDyn which incorporates the dynamics before confirming bond formation or breakage, allowing it to separate true chemical transformations from non-reactive encounters. This distinction is highlighted in Fig.~\ref{fig:R1_R2}(i), which shows a representative case. Here, the $\ce{H^{(1)}}-\ce{H^{(2)}}$ bond in $\ce{H_{2}}$ breaks, with $\ce{H^{(1)}}$ forming a stable $\ce{O-H^{(1)}}$ bond while $\ce{H^{(2)}}$ oscillates around the oxygen before departing, yielding the reaction $\ce{H_2 + O -> OH + H}$. ChemXDyn correctly identifies this transformation, whereas CTY-F and CTY-V misinterpret it as $\ce{H_2 + O -> H_2O}$, followed by $\ce{H_2O -> OH + O}$ in successive steps. This example demonstrates ChemXDyn’s ability to detect the true underlying chemistry with greater fidelity than existing methods. By contrast, panel (ii) in the figure shows a case where $\ce{H_2}$ and $\ce{O_2}$ approach each other and then separate without reacting. ChemXDyn properly recognizes this as a non-reactive collision, while CTY-F and CTY-V erroneously classify it as the reaction $\ce{H_2 + O_2 -> HO_2 + H}$, immediately followed by its reverse \textcolor{blue}{(see \textit{SI video V1})}. Such false positives, in which transient encounters are misidentified as reactions, inflate the apparent number of reaction events in CTY-F and CTY-V analyses (Fig.~\ref{fig:H2_results}(C)). This spurious detection artificially elevates computed rate constants, since the rate constant ($k$) is directly proportional to the number of occurrences (Eq.~\ref{eq:rate_constant}).

The impact of these methodological differences is evident in the Arrhenius-type $\ln k$ versus $10000/T$ plots used for rate constant evaluation (Fig.~\ref{fig:lnk_plots}). Each panel presents rate constants computed at different temperatures for representative reactions and compares them against the reference values from the Li \textit{et al.} mechanism \cite{li} for $\ce{H2}/\ce{O2}$ combustion. CTY-F and CTY-V systematically overpredict rate constants, often by more than an order of magnitude, owing to their tendency to misclassify non-reactive collisions as chemical events. This results in inflated slopes and offsets relative to the Li \textit{et al.} mechanism profiles. In contrast, ChemXDyn closely follows the reference trends across all temperatures, producing rate constants that lie within the expected range and preserve the correct Arrhenius behavior. These results highlight how ChemXDyn’s dynamics-based filtering and chemically informed bond assignment effectively suppress false positives, thereby enabling more accurate identification of reaction pathways and estimation of rate constants. Together, these capabilities provide a robust foundation for kinetic parameter extraction and their subsequent use in quantitative combustion modeling.

\subsubsection*{$\mathbf{{NH_3}/{N_2}/{O_2}}$ MD trajectory analysis}

Having demonstrated ChemXDyn’s performance for $\mathrm{H_2}/\mathrm{O_2}$ combustion, we now turn to reactive MD trajectories in the case of ammonia oxidation. Unlike the simpler diatomic fuel system ($\mathrm{H_2}/\mathrm{O_2}$), ammonia presents a far more complex chemistry, involving additional bond types (N–H, N–N, and N–O) and the formation of reactive intermediates that are challenging to resolve. In particular, BO distribution for the N–N bond exhibit three distinct peaks (see Fig.~\ref{fig:flowchart}(B)) corresponding to single, double, and triple bonds, while nitrogen atoms themselves can adopt multiple oxidation states. These features make the accurate detection of nitrogen chemistry difficult when relying on static BO threshold. In fact, conventional trajectory analyzers such as ChemTrYzer (CTY) have been reported to perform poorly for nitrogen-containing systems, to the extent that their applicability to such chemistry is explicitly cautioned against \cite{dontgen2015automated}. Nevertheless, CTY continues to be used in prior studies of ammonia combustion \cite{guo2023reactive, li2024chemical, wang2024role, ammonia-methane}, which could lead to inconsistent or misleading mechanistic interpretations. Therefore, this context makes ammonia a stringent test case for evaluating the robustness and chemical reliability of ChemXDyn. 

\begin{table}[!h]
\centering
\small
\begin{tabular}{@{}ccc@{}}
\toprule
\textbf{} & \textbf{Species count} & \textbf{Reaction count} \\
\midrule

CTY-F & 68 & 273 \\
ChemXDyn & 39 & 173 \\
Nakamura \textit{et al.} \cite{nakamura2017kinetic} & 20 & 112 \\

\bottomrule
\end{tabular}
\caption{Benchmarking species and reaction count for CTY-F and ChemXDyn results with $\ce{NH3}/\ce{N2}/\ce{O2}$ mechanism \cite{nakamura2017kinetic}.}
\label{tab:NH3_species_reactions}
\end{table}


For this analysis, we considered $NH_{3}/O_{2}/N_{2}$ system (see Case~3 in Table~\ref{tab:1}) and post-processed the atomistic trajectories using both CTY-F and ChemXDyn. The total number of species and reactions identified by each method are summarized in Table~\ref{tab:NH3_species_reactions} and benchmarked against the reference mechanism reported by Nakamura \textit{et al.} \cite{nakamura2017kinetic}. CTY-F detects 69 species and 273 reactions in total, whereas ChemXDyn identifies 39 and 173, respectively. In contrast, Nakamura \textit{et al.} explained the essential ammonia oxidation pathways with only about 20 key species and corresponding 112 reactions by employing sensitivity analysis. A notable observation from CTY-F is its inability to capture important reactions such as $\ce{NH2 + HO2 -> H2NO + OH}$, $\ce{N2O + H2O2 -> HN2O + HO2}$, $\ce{N2O + H -> N2 + OH}$ and $\ce{HO + N2 -> HN2 + O}$ which are critical reactions to evaluate emissions performance \cite{nakamura2017kinetic,klippenstein2022theoretical, glarborg2021rate, tamaoki2024roles}.
However, ChemXDyn successfully detects these reactions, yielding a more consistent representation of the reaction pathways. 

\begin{figure}[h!]
    \centering
    \includegraphics[width=1\linewidth]{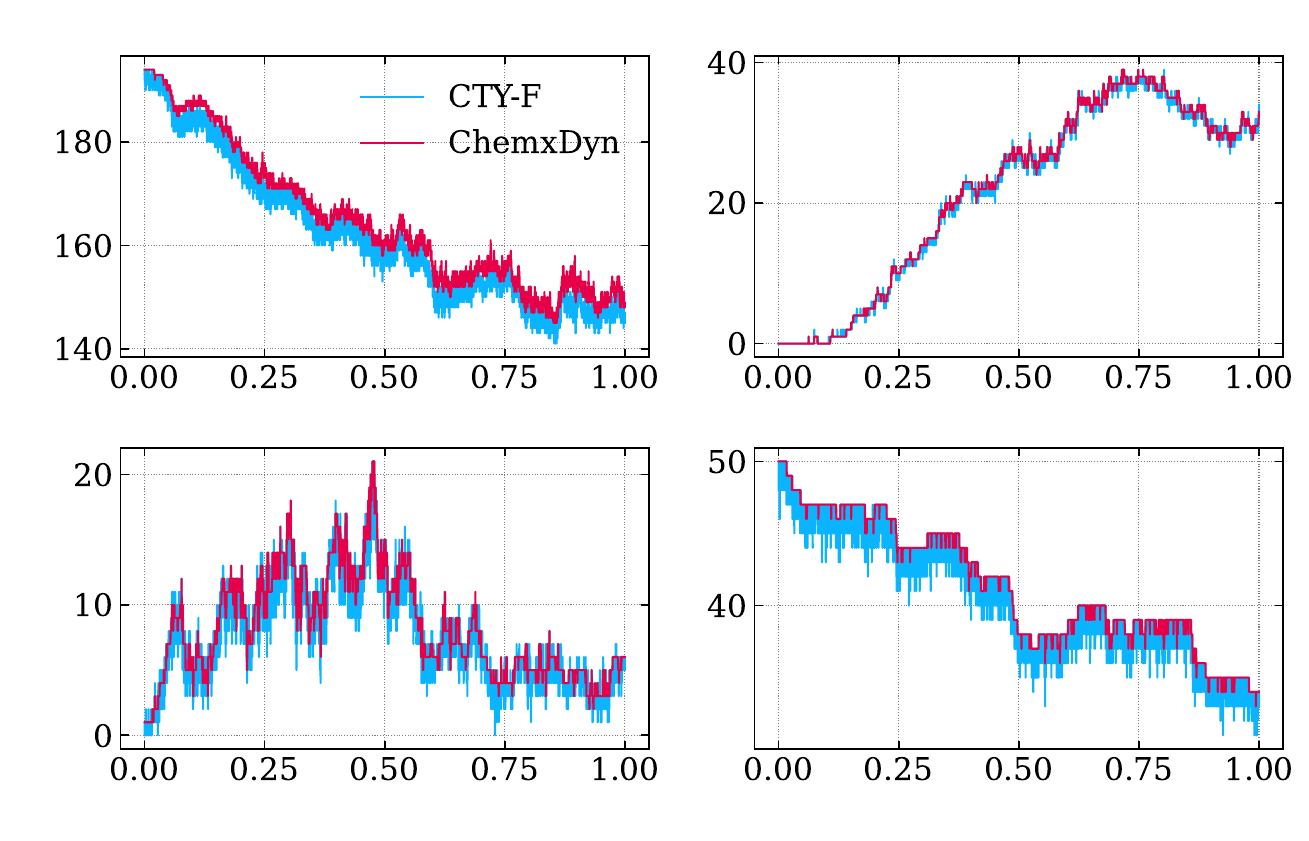}
    \caption{Time evolution of key species in the $\ce{NH3}/\ce{N2}/\ce{O2}$ oxidation system extracted using CTY-F (representative of CTY-V as well) and ChemXDyn.}
          \begin{tikzpicture}[remember picture,overlay]
      \node[rotate=90] at (-6.3,3.8) { $\ce{NH2}$ count};
      \node[rotate=90] at (-6.3,7.8) { $\ce{N2}$ count};
      \node[rotate=90] at (0.35,3.8) { $\ce{O2}$ count};
      \node[rotate=90] at (0.35,7.8) { $\ce{N2H2}$ count};

      \node[rotate=0] at (-2.6,5.8) {Time (ns)};
      \node[rotate=0] at (3.7,5.8) {Time (ns)};
      \node[rotate=0] at (-2.6,1.8) {Time (ns)};
      \node[rotate=0] at (3.7,1.8) {Time (ns)};
      
    \end{tikzpicture}
    \label{fig:NH3_time_evolution}
\end{figure}

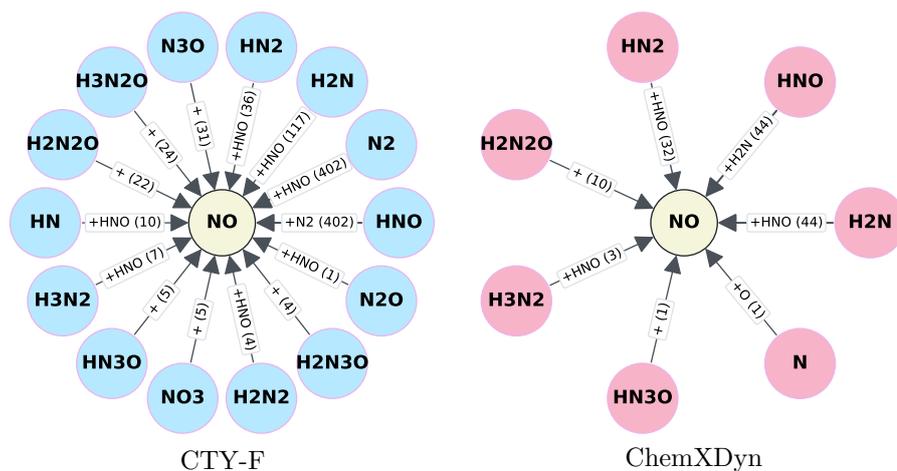
\begin{figure}[h!]
    \centering
  \includegraphics[width=.45\columnwidth]{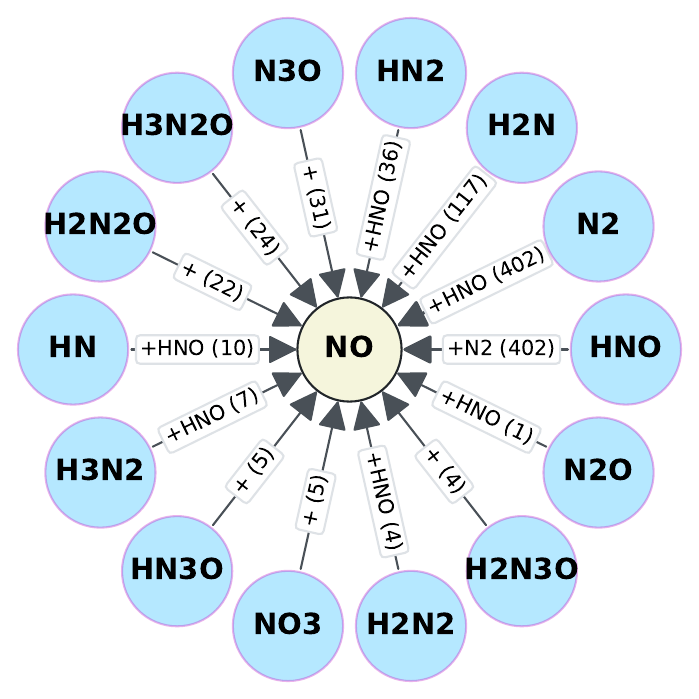}\hspace{2mm}
  \includegraphics[width=.45\columnwidth]{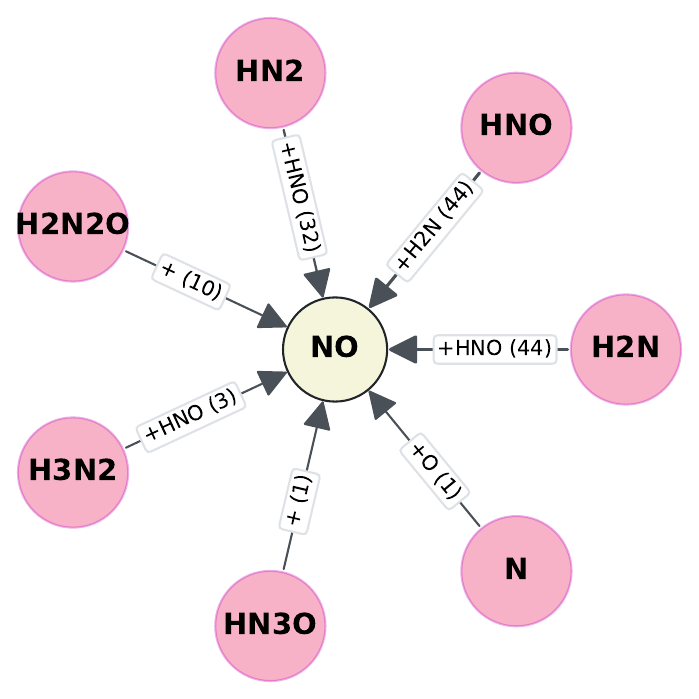}
  \begin{tikzpicture}[remember picture,overlay]

      \node[rotate=0] at (-9.25,-0.2) {CTY-F};

      \node[rotate=0] at (-3.05,-0.2) {ChemXDyn};
      
    \end{tikzpicture}

    \vspace{3mm}

  \caption{$\ce{NO}$-formation pathways extracted using CTY-F (sky blue) and ChemXDyn (light red). Outer nodes indicate $\ce{NO}$ precursors, and edge labels denote the other reactant involved in each formation pathway.}
  \label{fig:Nox}
\end{figure}

\begin{figure}[h!]
    \centering
  \includegraphics[width=.45\columnwidth]{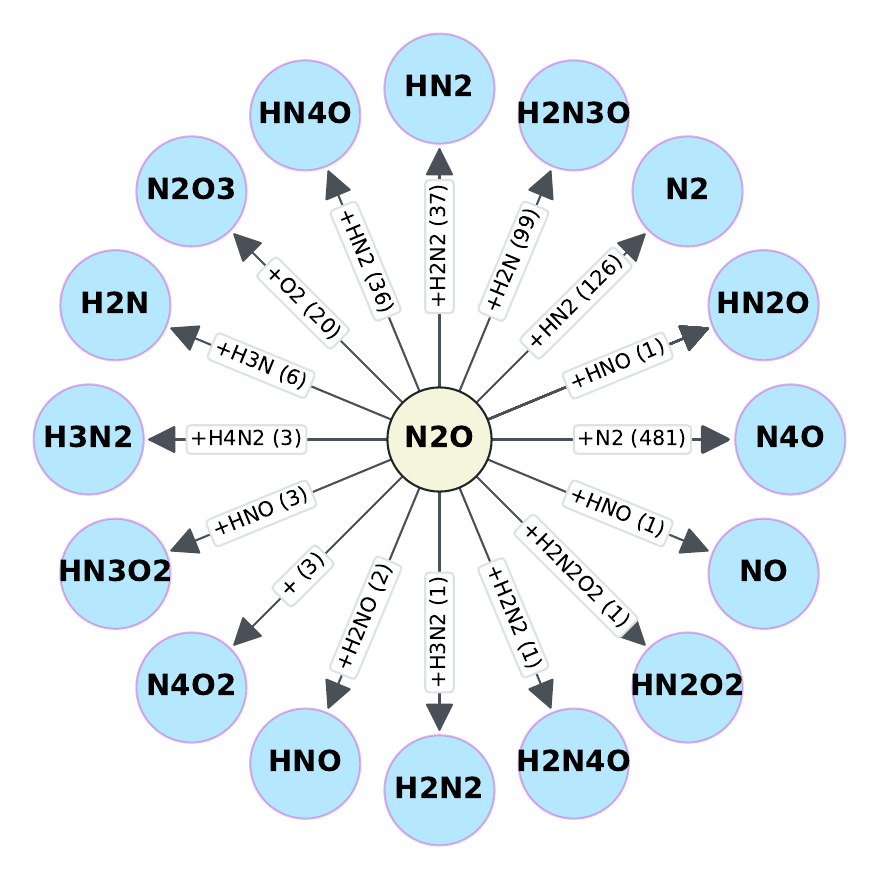}\hspace{2mm}
  \includegraphics[width=.45\columnwidth]{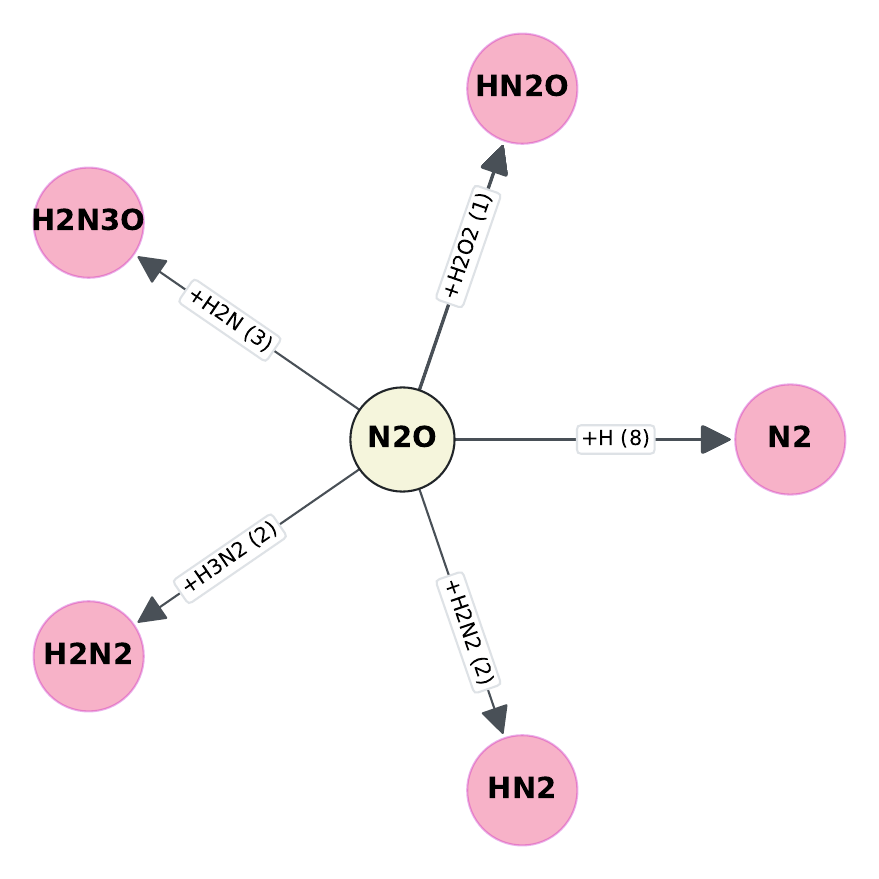}
  \begin{tikzpicture}[remember picture,overlay]

      \node[rotate=0] at (-9.25,-0.2) {CTY-F};

      \node[rotate=0] at (-3.05,-0.2) {ChemXDyn};
      
    \end{tikzpicture}

    \vspace{3mm}

  \caption{$\ce{N2O}$-consumption pathways extracted using CTY-F (sky blue) and ChemXDyn (light red). Outer nodes represent products formed after $\ce{N2O}$ consumption, and edge labels indicate the other reactant participating in each consumption pathway.}
  \label{fig:Nox_N2O}
\end{figure}

 To further illustrate these differences between CTY-F and ChemXDyn, we next examine the time evolution of selected species obtained in both these cases. Figure~\ref{fig:NH3_time_evolution} shows these evolution profiles for a subset of representative species, while other important species profiles are reported in \textcolor{blue}{(see \textit{SI Figure B3})}. Similar to the $\mathrm{H_2}/\mathrm{O_2}$ system, species in ammonia combustion exhibit pronounced fluctuations when analyzed using CTY-F. ChemXDyn, on the other hand, substantially reduces these spurious oscillations and yields smoother temporal profiles that better reflect the underlying dynamics.  Figure~\ref{fig:Nox} illustrates the sources of species contributing to $\ce{NO}$ formation. CTY-F predicts a much larger set of contributing species including several unphysical ones such as $\ce{N3O}$ than ChemXDyn, primarily due to its inability to distinguish between reactive and non-reactive collisions. Furthermore, examination of the $\ce{NO_x}$ chemistry further underscores these differences. For example, the reaction $\ce{NH2 + HO2 -> H2NO + OH}$, an important pathway in ammonia oxidation \cite{klippenstein2022theoretical, glarborg2021rate, tamaoki2024roles}, is captured by ChemXDyn but missed by CTY-F. Similarly, experimental studies and established mechanisms consistently highlight $\ce{N2O}$ consumption via $\ce{N2O + H -> N2 + OH}$  and $\ce{N2O + H2O2 -> HN2O + HO2}$ \cite{glarborg2025experimental}, which is again recovered by ChemXDyn but absent in CTY-F as mentioned earlier (see Fig.~\ref{fig:Nox_N2O}). Although a comprehensive mapping of all reaction pathways will be presented in future work, the present results already demonstrate ChemXDyn’s robustness and accuracy in resolving chemically meaningful transformations.

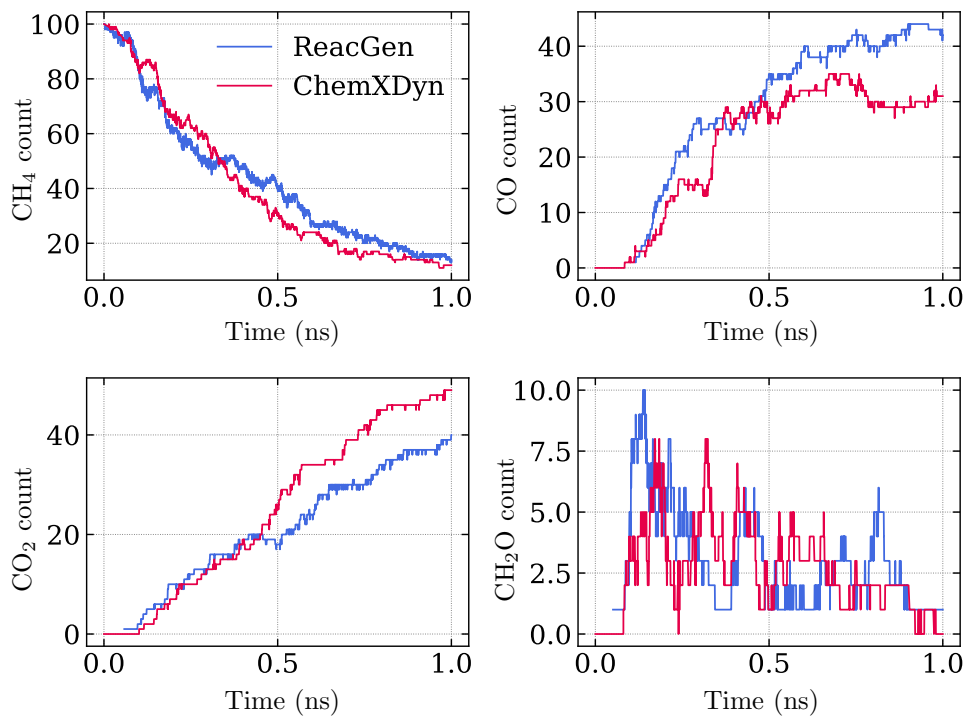
\begin{figure}[h!]
    \centering
    \includegraphics[width=1\linewidth]{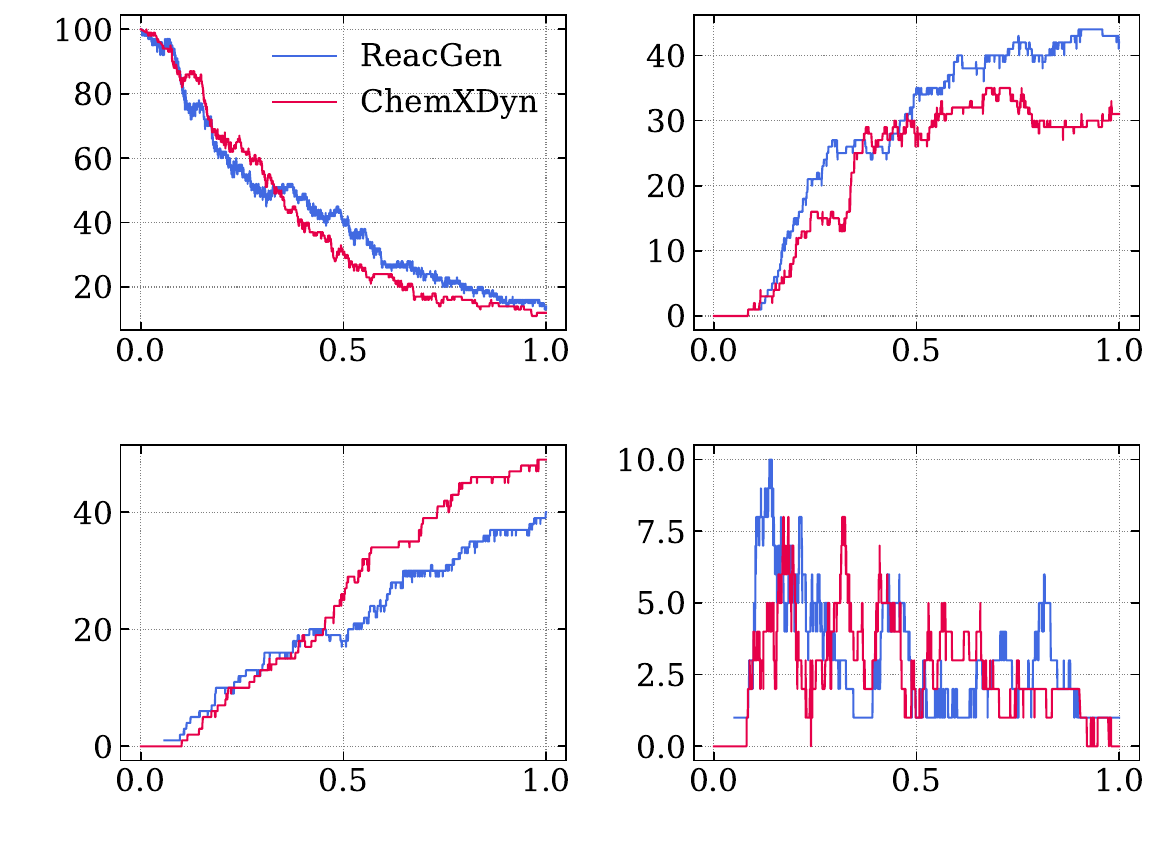}
   
          \begin{tikzpicture}[remember picture,overlay]
      \node[rotate=90] at (-6.0,3.0) {$\ce{CO2}$ count};
      \node[rotate=90] at (-6.0,7.9) {$\ce{CH4}$ count};
      \node[rotate=90] at (0.35,3.0) { $\ce{CH2O}$ count};
      \node[rotate=90] at (0.35,7.9) { $\ce{CO}$ count};

      \node[rotate=0] at (-2.6,5.6) {Time (ns)};
      \node[rotate=0] at (3.7,5.6) {Time (ns)};
      \node[rotate=0] at (-2.6,0.7) {Time (ns)};
      \node[rotate=0] at (3.7,0.7) {Time (ns)};
      
    \end{tikzpicture}

    \vspace{-5mm}

     \caption{Time evolution of key species in $\ce{CH4}/\ce{O2}$ NNMD trajectories analyzed using ReacNetGenerator (ReacGen) and ChemXDyn}
     
    \label{fig:CH4_time_evolution}
    
\end{figure}

 \begin{figure*}[h!]
 
    \centering
    \includegraphics[width=\linewidth]
    {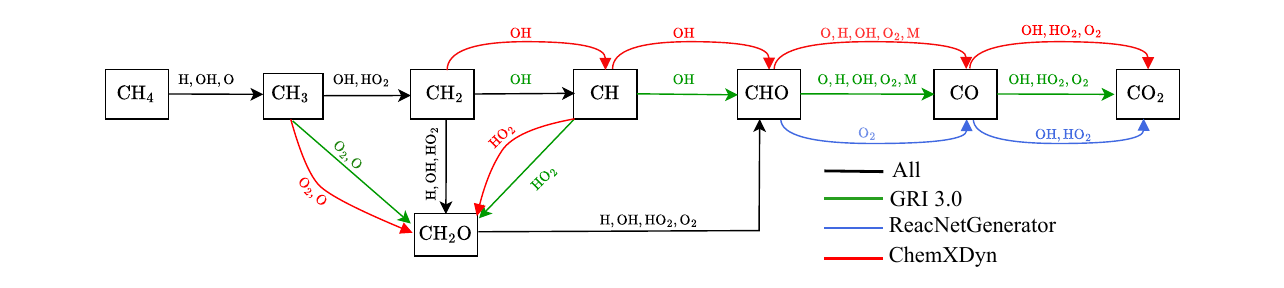}
    \caption{Comparison of dominant reaction pathways in $\ce{CH4}/\ce{O2}$ combustion extracted using ReacNetGenerator (purple) and ChemXDyn (red), validated against the GRI-Mech 3.0 mechanism (green). The pathways that are common to all are denoted using black arrows. ChemXDyn accurately reproduces the primary oxidation sequence consistent with GRI-Mech 3.0, while eliminating several spurious channels predicted by ReacNetGenerator.}
          \begin{tikzpicture}[remember picture,overlay]
    \end{tikzpicture}
    \label{fig:CH4_pathways}
\end{figure*}

\begin{figure}[!h]
    \centering
    \includegraphics[width=1\linewidth]{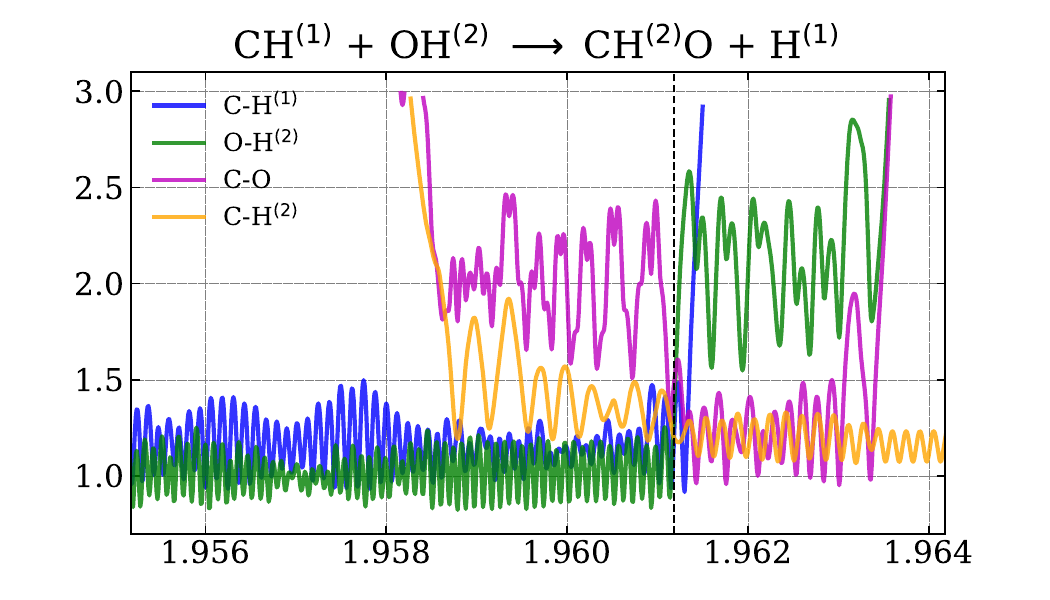}
    \caption{IAD vs. time plot illustrating the true occurrence of the reaction $\ce{CH + OH -> CHO + H}$ which demonstrates the importance of CH radical which ReacNetGenerator fails to detect.}
          \begin{tikzpicture}[remember picture,overlay]
      \node[rotate=90] at (-5.75,5.5) {IAD ($\mathrm{\AA}$)};

      \node[rotate=0] at (0.2,1.6) {Time (ns)};
      
    \end{tikzpicture}
    \label{fig:CH_reaction}
\end{figure}

\subsubsection*{$\mathbf{{CH_4}/{O_2}}$ MD trajectory analysis}
To further assess the versatility of ChemXDyn, we extended our analysis to the $\mathrm{CH_4}/\mathrm{O_2}$ combustion system simulated using NNMD (see Table~\ref{tab:1}). NNMD employs MLIP \cite{nnmd} trained on high-fidelity  \textit{ab initio} data, enabling accurate trajectory generation at reduced computational cost. Compared to the previously discussed systems, where BOs were employed to define chemical connectivity, this system was analyzed using interatomic distances (IADs). This allows for a direct evaluation of ChemXDyn’s robustness when applied to datasets that do not explicitly contain BO information. The $\mathrm{CH_4}/\mathrm{O_2}$ system, characterized by complex oxidation involving large number of radicals, thus serves as an ideal benchmark for testing the method’s generality under IAD-based analysis. To establish a fair comparison, ChemXDyn’s performance was evaluated against ReacNetGenerator \cite{reacnetgenerator}, which also relies on IADs for species identification but employs its own internal filtering criteria based on species lifetimes and occurrences. Under these comparable conditions, ChemXDyn identified 71 species and 441 reactions, whereas ReacNetGenerator yielded 110 species and 701 reactions. Figure~\ref{fig:CH4_time_evolution} presents the time evolution of selected species obtained from NNMD trajectories using both approaches. The temporal profiles derived from ReacNetGenerator deviate noticeably from the expected combustion trends, consistent with our earlier observations for $\mathrm{H_2}/\mathrm{O_2}$, as shown in Fig.~\ref{fig:H2_results}(B), where its reliance on instantaneous atomic positions led to inaccurate temporal evolution patterns. 

We next analyze the reaction pathways extracted from the $\mathrm{CH_4}/\mathrm{O_2}$ NNMD trajectories and compare them with the established GRI-Mech~3.0 mechanism \cite{smith1999gri}, as illustrated in Fig.~\ref{fig:CH4_pathways}. In the figure, notably, ReacNetGenerator fails to detect the $\ce{CH}$ radical, an essential intermediate in methane oxidation \cite{zettervall2021evaluation, Tonarely2022, Blevins1999} which leads to the omission of several key pathways. In contrast, ChemXDyn accurately reproduces the principal oxidation sequence of methane, encompassing the progression 
$\ce{CH4 \rightarrow CH3 \rightarrow CH2 \rightarrow CH \rightarrow CHO/CH2O \rightarrow CO \rightarrow CO2}$, in close agreement with GRI-Mech~3.0~\cite{smith1999gri,li2019comparative,zettervall2021evaluation,mohapatra2022adaptability,wang2024study}.
This stepwise radical-mediated transformation has been widely recognized in detailed kinetic analyses of methane oxidation,
where successive hydrogen abstractions and oxidation of intermediate species (\ce{CH3}, \ce{CH2}, and \ce{CH}) ultimately yield 
formaldehyde, carbon monoxide, and carbon dioxide as terminal oxidation products~\cite{Gardiner1984Combustion,baulch2005evaluated,crabtree1995aspects}.
The nitrogen-free subset of GRI-Mech~3.0 comprises 35 species and 217 reactions, of which ChemXDyn recovers 58 matching reactions compared to only 22 identified by ReacNetGenerator. Crucially, ChemXDyn captures the $\ce{CH}$-related chemistry, the $\ce{CHO}$ to  $\ce{CO}$ conversion through radical reactions with $\ce{H}$, $\ce{OH}$, and $\ce{O}$, as well as the three-body process $\ce{CHO + M -> CO + H + M}$, all of which are absent in ReacNetGenerator (see Fig.~\ref{fig:CH4_pathways}). As an illustration, Fig.~\ref{fig:CH_reaction} shows the $\ce{CH + OH -> CHO + H}$ reaction as a IAD–versus-time profile, accurately resolved by ChemXDyn. ReacNetGenerator fails to capture this process because its static threshold based representation merges $\ce{CH}$ and $\ce{OH}$ into a single species ($\ce{CH2O}$), thereby missing the transient radical intermediate. By correctly preserving such short-lived yet chemically crucial radicals, ChemXDyn reconstructs the full oxidation sequence and its branching pathways with high fidelity. This agreement with GRI-Mech~3.0 demonstrates that ChemXDyn reliably reconstructs chemically accurate reaction networks, even when applied to IAD-based NNMD data.

\section*{Discussion}

This work introduced ChemXDyn, a novel dynamics-based methodology for extracting chemically meaningful species and reaction pathways directly from molecular dynamics (MD) simulations. By explicitly incorporating temporal information into the analysis of interatomic distances or bond orders, ChemXDyn addresses the limitations of conventional distance-based approaches that rely solely on instantaneous atomic positions. Through the use of moving-window averaging, bond order distributions, and valence/coordination check, the methodology discriminates genuine bond formation and breakage events from transient vibrational fluctuations or non-reactive collisions: an essential capability for producing accurate kinetic models.

Comprehensive evaluation across three representative fuel-oxidation systems: hydrogen, ammonia, and methane, demonstrated ChemXDyn’s robustness and generality. For hydrogen oxidation, ChemXDyn accurately reproduced stable \ce{O2} dynamics, eliminated unphysical intermediates such as \ce{O4} and \ce{O6}, and yielded rate constants closer with the Li \textit{\textit{et al.}} mechanism. In ammonia-air oxidation, where conventional analyzers struggle due to multiple nitrogen oxidation states and additional N–H, N–N, and N–O bonding environments, ChemXDyn successfully captured key $\ce{NO_x}$/$\ce{N2O}$-forming and -consuming pathways, including \ce{NH2 + HO2 -> H2NO + OH} and \ce{N2O + H -> N2 + OH}, achieving close agreement with the experimental mechanism of Nakamura \textit{et al.} Extending the framework to methane oxidation using neural-network MD further confirmed its versatility: ChemXDyn recovered the full oxidation sequence $\ce{CH4 \rightarrow CH3 \rightarrow CH2 \rightarrow CH \rightarrow CHO/CH2O \rightarrow CO \rightarrow CO2}$ in strong agreement with the GRI-Mech 3.0 mechanism and experimental studies, while alternative distance-based analyses failed to detect transient radicals such as \ce{CH} and \ce{CHO}. Beyond reproducing known mechanisms, ChemXDyn provides a scalable and chemically interpretable bridge between atomistic simulations and macroscopic chemical kinetics. Its dynamic filtering, valence-aware bond assignment, and compatibility with both ReaxFF and machine learned interatomic potentials establish a transferable foundation for automated reaction discovery, mechanism validation, and rate constant estimation. The framework’s modular design enables seamless integration into high-throughput MD pipelines and data-driven combustion modeling workflows, extending its applicability to catalytic, plasma, and electrochemical systems where reaction dynamics are inherently complex. In summary, ChemXDyn transforms how reactive trajectories are interpreted, shifting from static structural snapshots to a temporally resolved, chemically faithful understanding of molecular reactivity. 

\section*{Simulation details}
In this study, three fuel oxidation systems with fuels being hydrogen ($\mathrm{H_2}$), ammonia ($\mathrm{NH_3}$), and methane ($\mathrm{CH_4}$), were investigated, as summarized in Table \ref{tab:1}. Each system was initialized within a cubic simulation box whose dimensions were adjusted to achieve a uniform density of 0.5 g cm$^{-3}$. Initial molecular configurations were generated using Packmol \cite{packmol}, ensuring uniform spatial distribution of species.
Reactive molecular dynamics (RMD) simulations were carried out using LAMMPS \cite{thompson2022lammps} with reactive force fields (ReaxFF) and machine learned interatomic potential (MLIP). ReaxFF parameter sets were chosen according to each system: C/H/O-2011 \cite{agrawalla2011development} for the $\mathrm{H_2/O_2}$ system and C/H/O/N-2019 \cite{kowalik2019atomistic} for $\mathrm{NH_3/N_2/O_2}$.
To complement these classical reactive potentials and further evaluate the generality of ChemXDyn, we additionally analyzed trajectories generated from a neural network-based MD (NNMD) using reference MLIP for the $\mathrm{CH_4/O_2}$ system. This MLIP, originally trained on high-fidelity \textit{ab initio} datasets (as reported in Ref.~\cite{nnmd}), was implemented within the LAMMPS framework to accurately reproduce short-range chemical reactivity and long-range intermolecular interactions. All simulations were conducted in the canonical  ensemble (NVT) for both equilibration and production stages, with temperature controlled via a Nosé–Hoover thermostat \cite{evans1985nose} and a damping parameter of 25fs. Periodic boundary conditions were applied in all three spatial directions, and an uniform integration timestep of 0.05fs ($\mathrm{H_2/O_2}$ system) and 0.1fs ($\mathrm{NH_3/N_2/O_2}$, $\mathrm{CH_4/O_2}$) were used. Initial atomic velocities were assigned from the Maxwell–Boltzmann distribution at 300K, followed by gradual heating to the target temperatures listed in Table \ref{tab:1}. To prevent premature reactions during heating, selected bond parameters (e.g., O–H in the $\mathrm{H_2/O_2}$ system) were temporarily deactivated. Once thermal equilibrium was reached, production runs were carried out under NVT ensemble. The resulting atomic trajectories from both ReaxFF-based RMD and MLIP-based NNMD simulations were subsequently analyzed using ChemXDyn to identify chemical species and their temporal evolutions, reactions and rate constants.

\bmhead{Supplementary information}


Supplementary information contains appendix A, B, figures A1, B2, B3, a table A1, two videos V1 and V2. 




\section*{Declarations}
The authors declare no competing interest.

K.A. and P.M. conceived the basic idea with inputs from E.A.; R.M. and D.B. developed and implemented the methodology, performed the computational studies and analyzed the results; K.A. and P.M. contributed to conceptualisation at every step, project administration, funding acquisition, supervision, validation, and manuscript review and editing. All authors contributed to writing the first draft of the manuscript.

KA was supported by the ANRF Core Research Grant, India. PM was supported by Google Research India Fellowship.








\newpage

\begin{appendices}



\section{Determination of equilibrium bond lengths and bond orders from IAD/BO distributions} \label{sec:supp1}
\begin{figure}[h!]
\centering
\includegraphics[width=\textwidth]{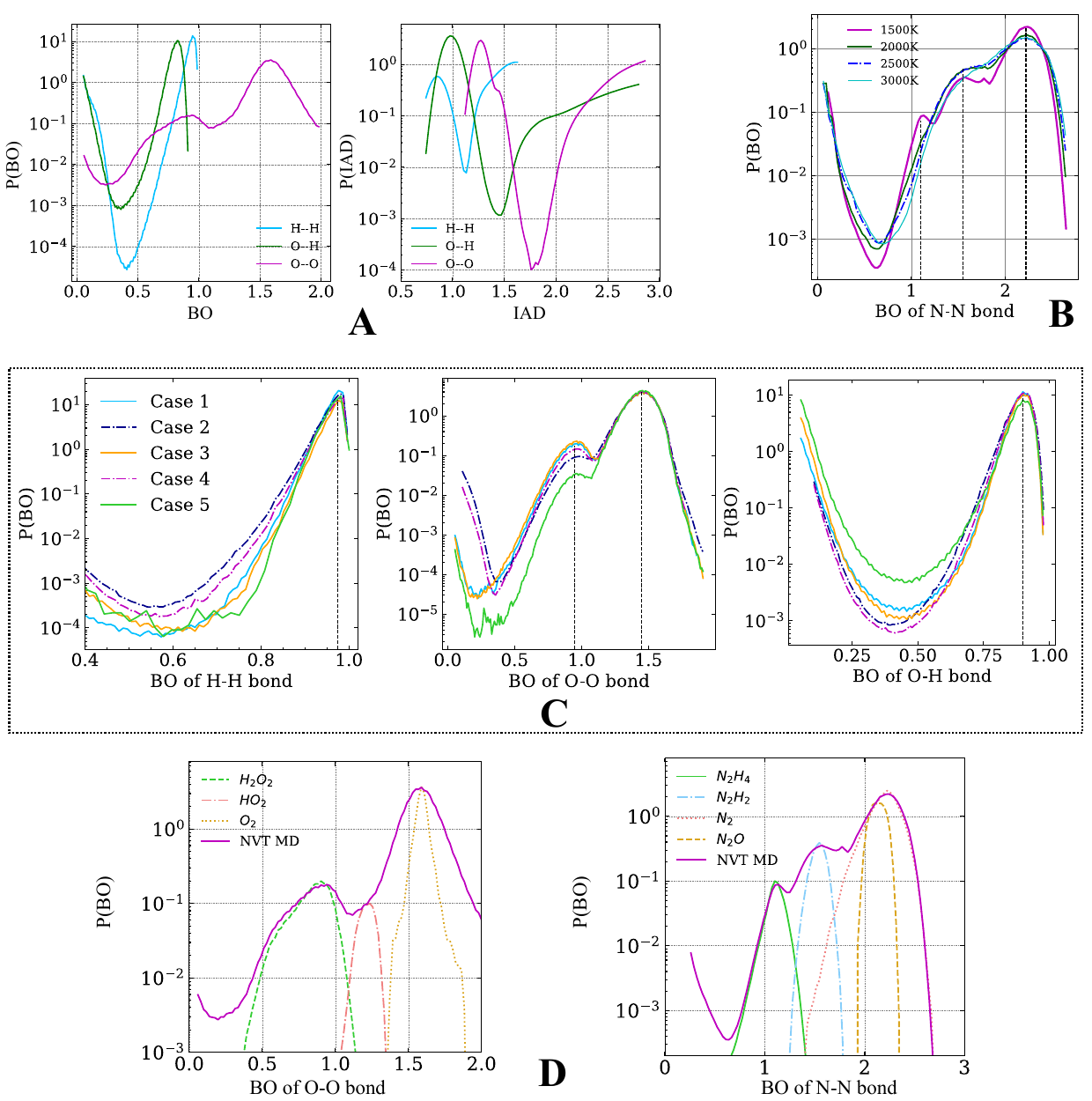}
\caption{Bond-order and interatomic-distance (BO/IAD) distributions illustrating bond multiplicity and equilibrium properties.
(A) BO and IAD distributions for bond types in the $\ce{H2}/\ce{O2}$ oxidation system, showing two peaks for O–O and single peaks for H–H and O–H bonds.
(B) Temperature-dependent BO distributions for N–N bonds from NVT simulations, where higher temperatures cause peak broadening and merging of bonded and non-bonded regions; at 1500 K, three distinct peaks correspond to single, double, and triple bonds.
(C) BO distributions of H–H, O–H, and O–O bonds across five simulation cases (Table \ref{tab:bo_dist_cases}), showing consistent equilibrium peak positions (black dashed lines) but varying minima with system conditions.
(D) Validation of BO-distribution peaks using independent MD simulations of $\ce{H2O2}$, $\ce{O2}$, and $\ce{HO2}$ (O–O bonds) and $\ce{N2H4}$, $\ce{N2H2}$, $\ce{N2}$, and $\ce{N2O}$ (N–N bonds), demonstrating that the peak positions align closely with those from the NVT MD trajectories of fuel oxidation.}
\label{fig:bo_plots_all}
\end{figure}

Molecular dynamics (MD) trajectories can be analyzed using either interatomic distances (IADs) or bond orders (BOs) to characterize the evolution of chemical bonding. In general, IAD provides a direct geometric measure of atomic separations, which can be used to identify bond formation and breaking events. However, in the case of ReaxFF-based MD, atomic interactions are parameterized explicitly in terms of BO, which serves as a continuous descriptor of bond strength derived from interatomic distance. The total potential energy of the system is evaluated as a function of these bond orders, making atomic connectivity and molecular topology entirely dependent on BO values. Consequently, understanding the BO–IAD relationship is essential for accurate species identification.
The total bond order between atoms $i$ and $j$ is composed of three components as given in equation:
\begin{align}
BO_{ij} &= BO^{\sigma}_{ij} + BO^{\pi}_{ij} + BO^{\pi\pi}_{ij} \label{eq:bo} \nonumber \\
        &= \exp \left[ p_{bo,1} \cdot \left( \frac{r_{ij}}{r^{\sigma}_o} \right)^{p_{bo,2}} \right]
         + \exp \left[ p_{bo,3} \cdot \left( \frac{r_{ij}}{r^{\pi}_o} \right)^{p_{bo,4}} \right]
        + \exp \left[ p_{bo,5} \cdot \left( \frac{r_{ij}}{r^{\pi\pi}_o} \right)^{p_{bo,6}} \right]
\end{align}
where $r^{\sigma}_o, r^{\pi}_o, r^{\pi\pi}_o$ represent equilibrium bond lengths of $\sigma$ bond, $\pi$ bond and double $\pi$ bond, respectively. $p_{bo}$ terms are empirical parameters to model the bonds. $BO_{ij}$ is the BO between atoms $i$ and $j$. $r_{ij}$ is the IAD between atoms $i$ and $j$ \cite{van2001reaxff}.
This equation describes how BO decays exponentially with increasing IAD, reflecting the weakening of atomic interactions as bonds stretch or break. To distinguish genuine chemical bonds from transient proximity events, we analyzed the BO/IAD distributions for all atom-pair types across the trajectory. For the $\ce{H2}/\ce{O2}$ system (see Table 1 in the main text), the corresponding distributions are shown in panels (i) and (ii) of Fig.~\ref{fig:bo_plots_all}(A). These distributions exhibit well-defined peaks corresponding to stable equilibrium bonding states (single, double, or triple bonds) and minima separating bonded and non-bonded regimes. The peak–valley pattern directly reflects the thermodynamic stability of these interactions energetically favored, long-lived bonds form sharp peaks, whereas transient or transition-state interactions populate the valleys between them.

Following the framework of Wu \textit{et al.}~\cite{jcc}, we interpret these distributions using the relation $\Delta G(x) = -RT \ln P(x)$, 
where $P(BO)$ is the BO distribution and $\Delta G(x)$ represents the associated free energy landscape. Minima in $P(BO)$ correspond to free energy maxima, effectively identifying the transition points between bonded and non-bonded states. Earlier studies \cite{jcc, reacnetgenerator,qi2012simulations} often adopted a BO threshold specific to atom-pair type (arrived from the first minima), neglecting the presence of multiple bonding modes. Our analysis reveals a more detailed picture: for example, while prior work \cite{jcc} identified only one prominent O–O peak under NVT conditions at 3500K, the present distributions, particularly at lower temperatures exhibit multiple distinct peaks corresponding to single, double, and even triple bonds. With increasing temperature, vibrational broadening (increase in amplitude) merges or flattens these peaks, making it progressively harder to distinguish between bonding states. This behavior is clearly visible in the N–N BO distributions for $\ce{NH3}$/$\ce{N2}$/$\ce{O2}$ systems at different temperatures, as shown in Fig.~\ref{fig:bo_plots_all}(B).

To examine this correspondence, we conducted five MD simulations spanning different fuel and thermochemical conditions (Table~\ref{tab:bo_dist_cases}). From these trajectories, the BO distributions of $\ce{H-H}$, $\ce{O-O}$, and $\ce{O-H}$ pairs which are common to all cases were extracted and are shown in Fig.~\ref{fig:bo_plots_all}(C). The equilibrium BO maxima remain remarkably consistent across compositions and temperatures, whereas the first minima separating bonded and non-bonded regimes shift significantly with temperature and system conditions. This systematic variation highlights a key limitation of using fixed or atom-pair-specific threshold values derived solely from these minima. The relationship between BO-distribution peaks and equilibrium BOs is further validated through independent MD simulations of $\ce{H2O2}$, $\ce{O2}$, and $\ce{HO2}$ (for O–O bonds) and $\ce{N2H4}$, $\ce{N2H2}$, $\ce{N2}$, and $\ce{N2O}$ (for N–N bonds), where the peak positions coincide closely with those observed in the NVT MD trajectories of fuel oxidation (Fig.~\ref{fig:bo_plots_all}(D)). Importantly, the same analytical framework can be extended to IAD-based MD simulations (e.g., neural-network or \textit{ab initio} MD), where distance trajectories are directly employed to extract equivalent information on bond stability and reactivity. Collectively, these analyses confirm that BO-distribution peaks provide a transferable and physically grounded reference for determining equilibrium BOs across systems and thermodynamic conditions, forming the foundation for robust bond detection in ChemXDyn.

\begin{table}\centering
\caption{Cases considered for equilibrium BO convergence study as shown in Fig.~\ref{fig:bo_plots_all}(C)}
  \centering
  \begin{tabular}{ccccc}
  \hline
    \textbf{Case} &  \textbf{Fuel}  &  \textbf{Number of Fuel/O$_{2}$/N$_{2}$}   &\textbf{Ensemble}  & \textbf{$T_{0}$ (K)} \\
    \hline
    \hline
    \textcolor{SkyBlue}{1} & \textcolor{SkyBlue}{H$_{2}$} & \textcolor{SkyBlue}{160/80/0} & \textcolor{SkyBlue}{NVT} & \textcolor{SkyBlue}{2500} \\
    \textcolor{blue}{2} & \textcolor{blue}{H$_{2}$} & \textcolor{blue}{100/50/195} & \textcolor{blue}{NVE} & \textcolor{blue}{1200} \\
    \hline
    \textcolor{orange}{3} & \textcolor{orange}{NH$_{3}$} & \textcolor{orange}{67/50/195} & \textcolor{orange}{NVT} & \textcolor{orange}{2500} \\
    \textcolor{magenta}{4} & \textcolor{magenta}{NH$_{3}$} & \textcolor{magenta}{67/50/195} & \textcolor{magenta}{NVE} & \textcolor{magenta}{1600} \\
    
    \hline
    
    \textcolor{LimeGreen}{5} & \textcolor{LimeGreen}{C$_{2}$H$_{2}$} & \textcolor{LimeGreen}{40/50/195} & \textcolor{LimeGreen}{NVT} & \textcolor{LimeGreen}{2000} \\
  

    \hline
    \vspace{1mm}
  \end{tabular}
  
  \label{tab:bo_dist_cases}
\end{table}

\section{Sensitivity of single-bond threshold and window size parameters}\label{sec:s3}
To ensure robust and chemically meaningful bond identification within ChemXDyn, we evaluated the sensitivity of two key parameters:
\subsection{Weak interaction threshold}
The weak-interaction threshold specifies the minimum fraction of the equilibrium bond order (or maximum for IAD) required for an interaction to be classified as a genuine chemical bond. Panel (i) of Fig.~\ref{fig:H2_parameter} displays three distinct maxima in the N–N bond order distribution, corresponding to single, double, and triple bonds. When the instantaneous bond order lies between any of the two maxima, the bond multiplicity is assigned based on the nearest peak. If the instantaneous bond order falls below the primary single-bond peak in the BO distribution, the weak-interaction threshold (red dashed line in Fig. \ref{fig:H2_parameter}) serves to distinguish true chemical bonds from transient, non-bonded encounters. Panel (ii) of Fig. \ref{fig:H2_parameter} illustrates the influence of this parameter on the time evolution of $\ce{H2}$ molecules. A very small threshold ($10 \%$ of the equilibrium value) leads to artificial bond breaking and recombination, resulting in pronounced fluctuations. In contrast, thresholds of $20 \%$ and $30 \%$ yield stable and nearly identical temporal profiles, indicating convergence. Based on this observation, a $20 \%$ single-bond threshold was adopted in all subsequent analyses.

\subsection{Window size}
\begin{figure}[h!]
\centering
    
\includegraphics[width=\textwidth]{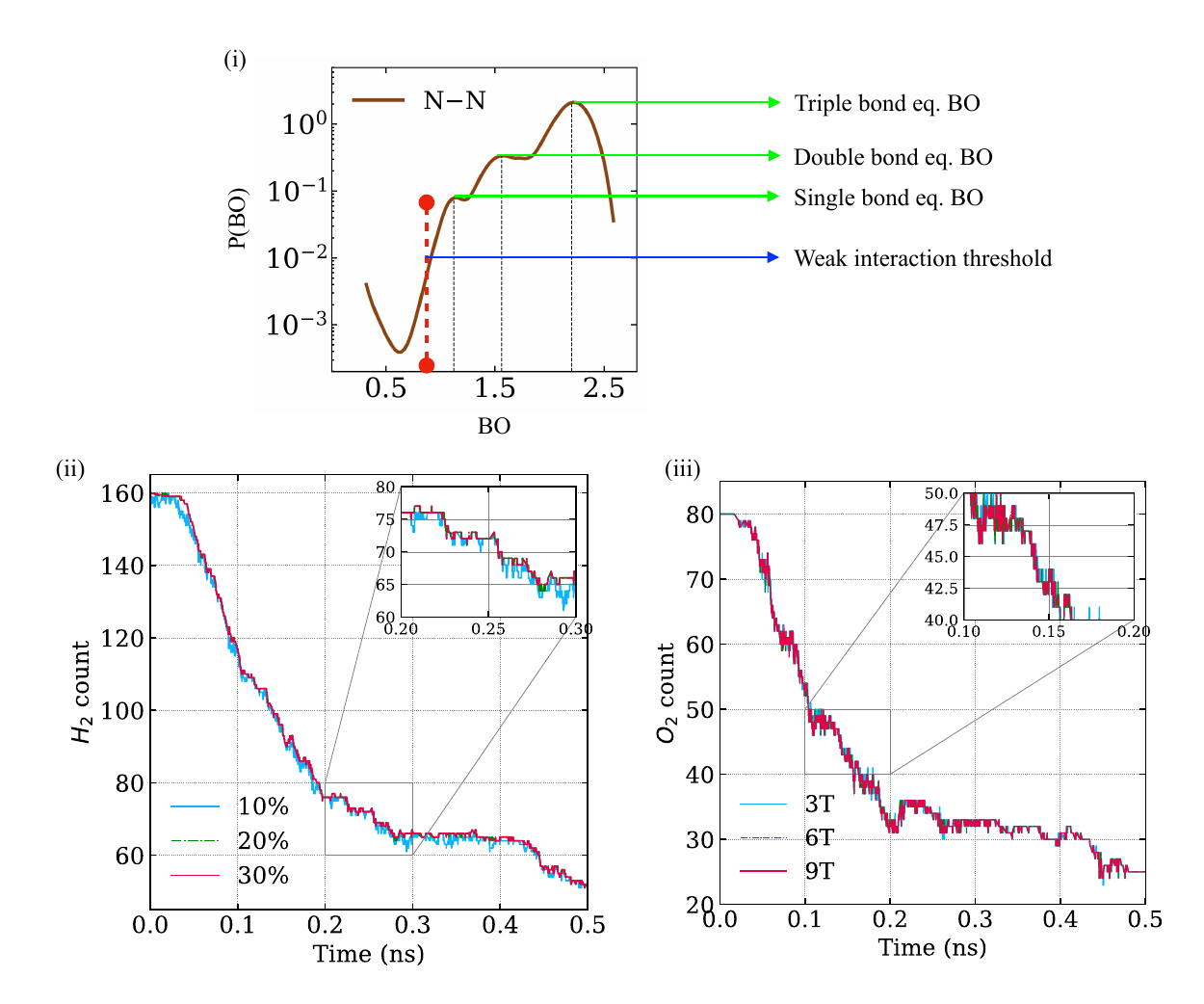}

  \caption{Sensitivity of single-bond threshold and window-size parameters in ChemXDyn. Panel (i) showing N-N BO distribution showcasing different multiplicities equilibrium BO values and also weak interaction threshold. Panel (ii) shows the number of $\ce{H2}$ molecules as a function of simulation time for weak interaction thresholds of $10 \%$, $20 \%$, and $30 \%$ of the equilibrium bond order value. Panel (iii) presents the $\ce{O2}$ time evolution for window sizes of $3 T$, $6 T$, and $9 T$, where $T$ denotes the fundamental $\ce{O2}$ vibrational period.}
  \label{fig:H2_parameter}
\end{figure}

The window-size parameter controls the length of the left (forward) and right (backward) averaging windows used to compute the mean bond order over time. Panel (iii) of Fig. \ref{fig:H2_parameter} shows the effect of varying the window size ($3 T$, $6 T$, and $9 T$), where $T$ represents the fundamental vibrational period of $\ce{O2}$ obtained from experimental data. The smallest window ($3 T$) introduces high-frequency fluctuations due to inadequate temporal averaging, while larger windows ($6 T$ and $9 T$) produce smoother, converged trajectories. Accordingly, a 6 T window size was chosen for all production simulations to balance temporal resolution and stability.

\begin{figure}[h!]
\centering
    
  \includegraphics[width=1\columnwidth]{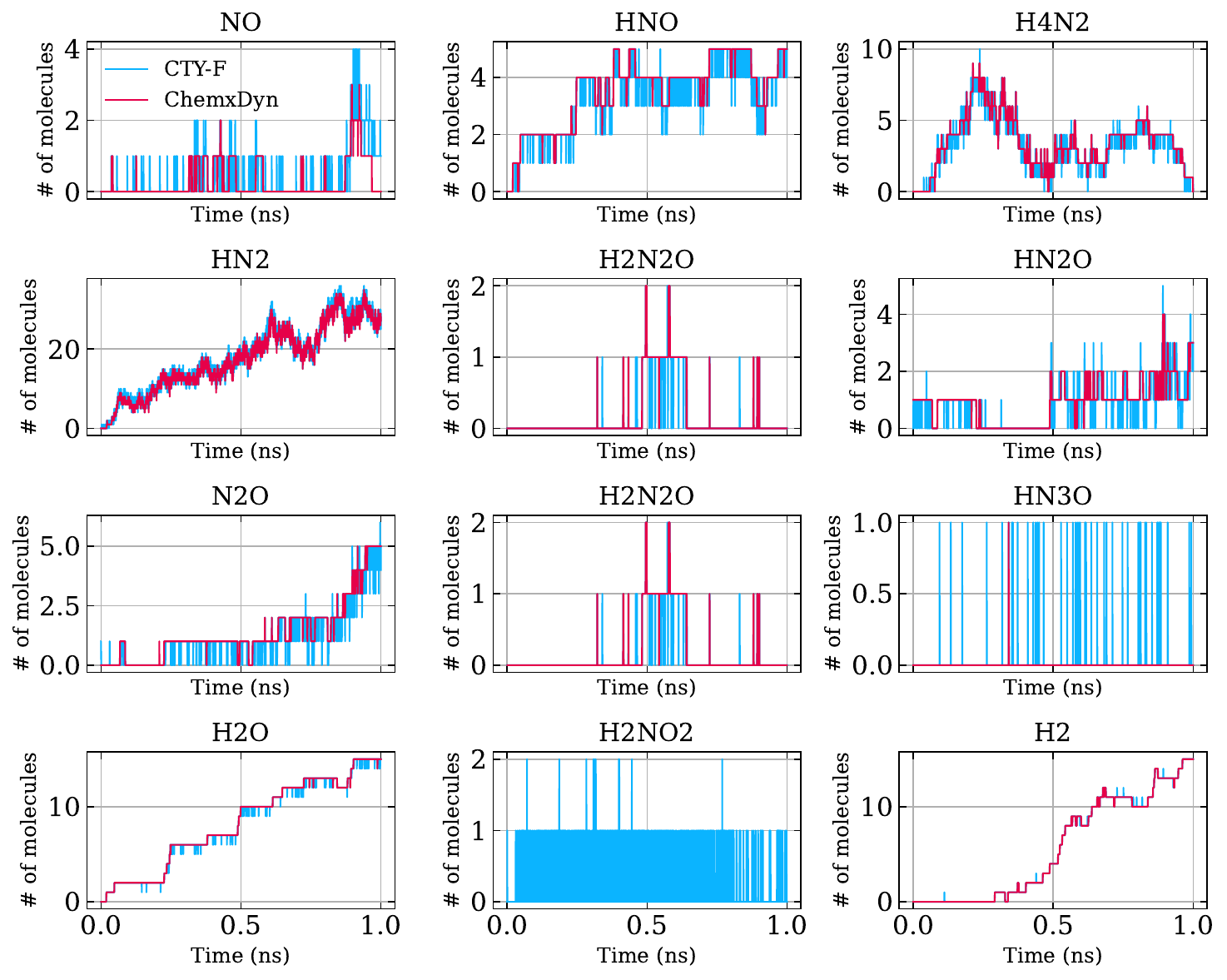}

      

  \caption{Time evolution of key species in the $\ce{NH3}/\ce{N2}/\ce{O2}$ oxidation system extracted using CTY-F (representative of CTY-V as well) and ChemXDyn.}
  \label{fig:NH3_species}
\end{figure}

\end{appendices}


\bibliography{sn-bibliography}

\end{document}